\documentclass[aps,prl,twocolumn,citeautoscript,nofootinbib]{revtex4-1}  
\usepackage{amsmath,amssymb,bm,bbm} 
\usepackage{graphicx}  
\usepackage{color} 
\usepackage[colorlinks=true]{hyperref}  
\hypersetup{
    bookmarks=true,         
    unicode=false,          
    pdftoolbar=true,        
    pdfmenubar=true,        
    pdffitwindow=false,     
    pdfstartview={FitH},    
    pdftitle={Theory of a Planckian Metal},    
    pdfauthor={Aavishkar Patel and Subir Sachdev},     
    pdfsubject={},   
    pdfcreator={},   
    pdfproducer={}, 
    pdfkeywords={} {} {}, 
    pdfnewwindow=true,      
    colorlinks=true,       
    linkcolor=magenta, 
    citecolor=blue,        
    filecolor=magenta,      
    urlcolor=blue           
} 

\usepackage{amsfonts}
\usepackage{upgreek}
\usepackage{slashed}
\usepackage{latexsym}
\usepackage{dsfont}
\usepackage{todonotes}

\newcommand{\beq}{\begin{equation}}
\newcommand{\eeq}{\end{equation}}
\def\bea{\begin{eqnarray}}
\def\eea{\end{eqnarray}}

\newcommand{\nn}{\nonumber \\}

\usepackage{braket}

\begin{document}

\title{Theory of a Planckian metal}

\author{Aavishkar A. Patel}
\affiliation{Department of Physics, Harvard University, Cambridge MA 02138, USA}

\author{Subir Sachdev}
\affiliation{Department of Physics, Harvard University, Cambridge MA 02138, USA}

\date{\today}

\begin{abstract}
We present a lattice model of fermions with $N$ flavors and random interactions which describes a Planckian metal at low temperatures, $T \rightarrow 0$, in the solvable limit of large $N$. 
We begin with quasiparticles around a Fermi surface with effective mass $m^\ast$, and then include random interactions which lead to fermion spectral functions with frequency scaling with $k_B T/\hbar$.
The resistivity, $\rho$,  obeys the Drude formula $\rho = m^\ast/(n e^2 \tau_{\textrm{tr}})$, where $n$ is the density of fermions, 
and the transport scattering  rate is $1/\tau_{\textrm{tr}} = f \, k_B T/\hbar$; we find $f$ of order unity, and essentially independent of the strength and form of the interactions. The random interactions are a generalization of the Sachdev-Ye-Kitaev models; it is assumed that processes non-resonant in the bare quasiparticle energies only renormalize $m^\ast$, while resonant processes are shown to produce the Planckian behavior.  
\end{abstract}
\maketitle

Electronic transport in metals has been successfully described by the Drude formula for many decades:
\beq
\rho = \frac{m^\ast}{n e^2} \, \frac{1}{\tau_\mathrm{tr}} \,, \label{drude}
\eeq
where $\rho$ is the resitivity, $n$ is the density of electrons, and $m^\ast$ is the effective mass of the quasiparticles at the Fermi surface.
The transport scattering time $\tau_{\textrm{tr}}$ is estimated from the quantum Boltzmann equation to obey
\beq
\frac{1}{\tau_\textrm{tr}} \sim U^2 T^2\,, \label{boltzmann}
\eeq
at low temperatures, $T$, where $U$ measures the strength of the electron-electron interactions. 

It was recognized early \cite{Takagi92} that the normal state of the cuprate high temperature superconductors does not obey the above paradigm, with a $T$-linear resistivity above the superconducting critical temperatures near optimal doping. This `strange' metal behavior has since been extensively studied, and has been found to be ubiquitous in correlated electron superconductors \cite{Taillefer10}. 
More recently, the $T$-linear resistivity has been quantitatively characterized \cite{MacKenzie13} by the Drude formula (\ref{drude}), using a value of $m^\ast$ in a nearby quasiparticle regime. A `Planckian' metal behavior has been 
observed \cite{Legros18, Paglione19,Pablo19,Bakr19}, with 
\beq
\frac{1}{\tau_\textrm{tr}} = f \frac{k_B T}{\hbar} \label{planck}
\eeq
with $f \approx 1 $ in correlated electron systems, twisted bilayer graphene, and ultracold atoms, where `Planckian' emphasizes the universality of Eq.~(\ref{planck}), dependent only on Planck's constant and the absolute temperature in units of energy. 
Specifically \cite{Legros18}, many cuprates and 
organic superconductors obey (\ref{planck}) at the lowest $T$, after suppression of superconductivity by a magnetic field, with $0.7 \lesssim  f \lesssim 1.2$.  This is in striking contrast to (\ref{boltzmann}), both in the $T$-linear dependence, and in the independence on the interaction strength $U$. Note that the value of $U$ varies over several orders of magnitude across the experimental systems noted above.

This letter will present a model of fermions with $N$ flavors and random interactions, which is solvable in the large-$N$ limit and whose transport realizes a Planckian metal. Our model is a lattice extension \cite{PG98,Zhang2017,Balents2017,Patel2017,Chowdhury2018,PatelKim2018} of the Sachdev-Ye-Kitaev (SYK) models \cite{SY92,kitaev2015talk}. Its ingredients are fermionic quasiparticles hopping on a lattice without disorder with dispersion $\epsilon_k$ near a Fermi surface, where $k$ is the crystal momentum. These quasiparticles have random interactions whose effective strength is weaker than the Fermi energy (see below (\ref{SYK2})), 
unlike previous lattice extensions \cite{Zhang2017,Balents2017,Patel2017,Chowdhury2018,PatelKim2018}. These interactions produce a Planckian metal state without well-defined quasiparticle excitations down to $T = 0$. A key feature of our model is that it explicitly retains only quasiparticle interactions which are resonant {\it i.e.} scattering of quasiparticles with momenta $k_1$ and $k_2$ to $k_3$ and $k_4$ is restricted to those obeying $\epsilon_{k_1} + \epsilon_{k_2} = \epsilon_{k_3} + \epsilon_{k_4}$. It is assumed that non-resonant interactions have already been accounted for by a suitable renormalization procedure, and absorbed into the quasiparticle dispersion $\epsilon_k$. Note that this `resonant selection' is closely analogous to that appearing in a renormalization group approach to Fermi liquid theory \cite{shankar_rg}: the Landau interactions $F^{a,s}_\ell$ act only on quasiparticles exactly on the Fermi surface, and so all interaction corrections are implicitly resonant; off-Fermi surface processes are assumed to have been absorbed into the values of $F^{a,s}_\ell$. We will show that a resonant lattice SYK model realizes a Planckian metal with resistivity 
obeying (\ref{drude}) and (\ref{planck}), with $f$ essentially independent of the strength and form of interactions and of order unity.

\noindent
{\bf The Model.} We consider quasiparticles annihilated by fermionic operator $c_{k \alpha}$, where $\alpha = 1 \ldots N$ is the SYK flavor index, with Hamitonian $H=H_k + H_U$
\bea
H_k &=& \int_k \sum_{\alpha} \epsilon_k c_{k \alpha}^\dagger c_{k\alpha}^{\vphantom\dagger}, \label{ressyk} \\
H_U &=&  \frac{1}{(2N)^{3/2}} \sum_{\alpha_a}
\int_{k_a} U_{\alpha_a} (k_a) c_{k_1 \alpha_1}^\dagger c_{k_2 \alpha_2}^\dagger c_{k_3 \alpha_3}^{\vphantom\dagger}
c_{k_4 \alpha_4}^{\vphantom\dagger}. \nonumber
\eea
Here $\int_k \equiv \int d^dk/(2\pi)^d$ in $d$ spatial dimensions. The $U_{\alpha_a} (k_a)$ (with $a = 1 \ldots 4$) are Gaussian random interactions with zero mean, whose second moment factorizes into flavor and momentum dependent factors
\beq
\overline{U_{\alpha_a}^\ast (k_a) U_{\alpha_{a'}'} (k_{a'}')} = \mathcal{U}_{\alpha_a \alpha_{a'}'}\,  \mathcal{K} (k_a, k_{a'}'). \label{UK}
\eeq
The $\mathcal{U}_{\alpha_a \alpha_{a'}'}$ are the same as those in the complex SYK model \cite{SS15}: they are non-vanishing, with values $\pm U^2$, only if 
all $\alpha_a$ equal $\alpha_{a'}'$, up to antisymmetrizations and Hermiticity implied by $H_U$ {\it e.g.} $U_{\alpha_1,\alpha_2,\alpha_3,\alpha_4} (k_1, k_2, k_3, k_4) = - 
U_{\alpha_2,\alpha_1,\alpha_3,\alpha_4} (k_2, k_1, k_3, k_4)$, $U_{\alpha_1,\alpha_2,\alpha_3,\alpha_4} (k_1, k_2, k_3, k_4) = U^\ast_{\alpha_4,\alpha_3,\alpha_2,\alpha_1} (k_4, k_3, k_2, k_1)$.

Without the $k$ index, $H$ as defined above is the well-studied `single-island' complex SYK model in which all quasiparticles have the same bare energy $\epsilon$. Such a model realizes a critical solution \cite{SY92,GPS01,SS15,Fu:2016yrv,Azeyanagi2018,GKST19} for a range of values of $|\epsilon| \lesssim 0.24 U$ \cite{Azeyanagi2018}, with a Green's function which is a universal scaling function of $\hbar \omega/(k_B T)$ ($\omega$ is the frequency) and a low energy spectral asymmetry parameter $\mathcal{E}$ determined by $\epsilon/U$: this scaling function will be described below. For $|\epsilon| \gtrsim 0.24 U$, there are no non-trivial solutions for the single island SYK model, and so the number density on the island for the non-quasiparticle state cannot be too far from half-filling \cite{Fu:2016yrv,Azeyanagi2018}; this feature of the SYK model will lead to some artifacts below in our model of the Planckian metal.

Including the $k$ dependence of the quasiparticle energy $\epsilon_k$, the simplest choice for the interactions is to have $\mathcal{K}$ momentum independent, apart from a momentum conserving delta function needed to preserve average translational invariance
\beq
\mathcal{K}_0 (k_a, k_{a'}') = \delta(k_1 + k_2 - k_3 - k_4 - k_1' - k_2' + k_3' + k_4'). \label{Kk0}
\eeq
In real space, (\ref{Kk0}) implies that the interactions are independent random variables on each island (associated with each lattice site), but with a site-independent variance. Such lattice SYK models have been thoroughly studied in recent work \cite{Zhang2017,Balents2017,Chowdhury2018,Patel2017,PatelKim2018}, and related models in earlier work \cite{PG98} (there is also a connection to the holographic analysis of a pair of SYK islands with single fermion hopping between them \cite{Maldacena:2018lmt}). Assuming the dispersion $\epsilon_k$ has bandwidth $W$, such a model realizes a non-Fermi liquid `incoherent metal' with the momentum-independent, critical Green's function of the single site model in the intermediate temperature regime $W^2/U \ll T \ll U$; a heavy Fermi liquid appears for $T \ll W^2/U$. The resistivity of the non-Fermi liquid is (in two spatial dimensions) $\rho \sim (h/e^2) (T/(W^2 / U))$. These properties are generic for almost all translationally invariant choices for $\mathcal{K}$, and have some drawbacks in comparison to the observed Planckian metals:
({\it i\/}) there is no sign of a Fermi surface in the non-Fermi liquid regime, which has been completely wiped out by the random interactions;
({\it ii\/})  the $T$-linear resistivity is always in a `bad metal' regime, obeying $\rho \gg h/e^2$, in contrast to observations with smaller resistivities;
({\it iii\/}) the co-efficient of the $T$-linear resistivity is strongly dependent upon $U$.

In passing, we note that recent work \cite{Chowdhury2018,Patel2017} 
also considered 2-band lattice generalizations of the SYK model, with a Kondo exchange interaction between itinerant electrons and SYK islands. These provide an explicit realization of a `marginal Fermi liquid' of the itinerant electrons, with properties similar to those in early 
studies \cite{Varma89,SY92,Faulkner13}. While such a marginal Fermi liquid is not a bad metal, it has other drawbacks in comparison to observations: ({\it i\/}) the Fermi surface is `small' and counts only the itinerant electrons, in contrast to the observed large Fermi surface which counts all electrons; ({\it ii\/}) the scattering rate $1/\tau_{\mathrm{tr}}$ is linear in $T$, but the prefactor is not universal and strongly dependent upon the strength of the Kondo exchange interaction.

The main point of the present letter is that a `resonant' choice for $\mathcal{K}$ leads to Planckian behavior down to $T \rightarrow 0$, 
without the drawbacks described above. The renormalization group rationale for such a choice was described above, in that the non-resonant interactions are accounted for by a renormalization of $\epsilon_k$. So we choose
\bea
\mathcal{K} (k_a, k_{a'}') &=& \mathcal{K}_0 (k_a, k_{a'}')\, \frac{1}{2} \Bigl[\mathcal{K}_1 (k_a) \delta(\epsilon_{k_1} + \epsilon_{k_2} - \epsilon_{k_3} - \epsilon_{k_4}) \nonumber \\
&+& \mathcal{K}_1 (k'_{a'}) \delta( \epsilon_{k_1'} + \epsilon_{k_2'} - \epsilon_{k_3'} - \epsilon_{k_4'}) \Bigr], \label{Kk}
\eea
where $\mathcal{K}_1$ is a smooth function of momenta. We will assume inversion symmetry upon disorder average in our system, so $\epsilon_{k}=\epsilon_{-k}$ and $\mathcal{K}_1(k_a)=\mathcal{K}_1(-k_a)$. The additional delta function momentum dependence in (\ref{Kk}), over that in (\ref{Kk0}), implies long-range power-law correlations in the random interactions in real space. 

\begin{figure}
    \centering
    \includegraphics[width=0.48\textwidth]{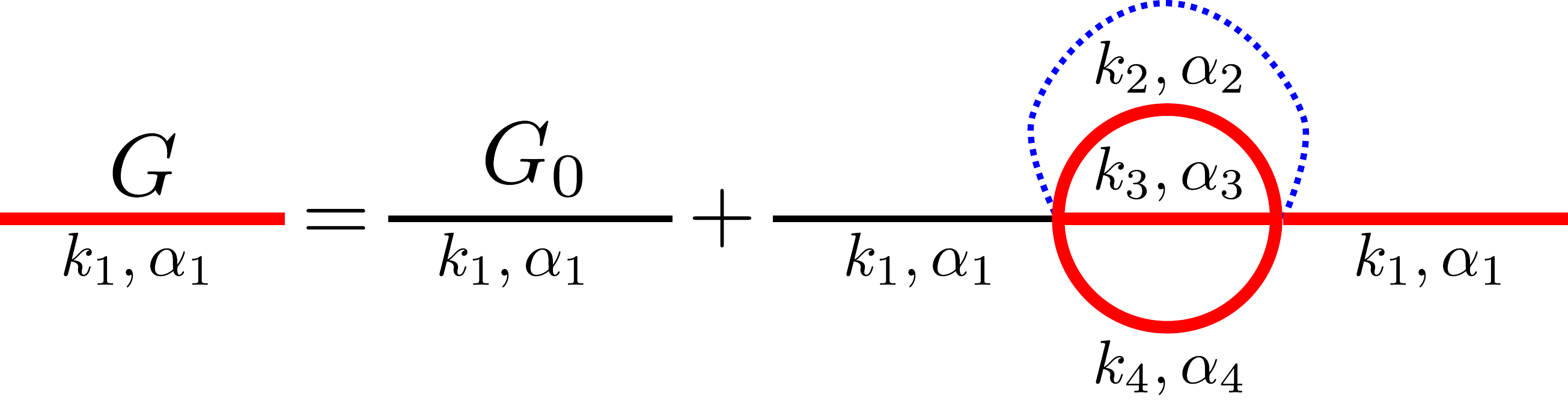}
    \caption{Self-consistent Dyson equation, exact in the large-$N$ limit, for the fermion propagator $G(k,\omega)=(1/N)\sum_{\alpha=1}^N\langle c_{k,\alpha}(\omega)c^\dagger_{k,\alpha}(\omega)\rangle$ (thick red line). The dashed blue line denotes the disorder averaging of the random couplings $U_{\alpha_a}(k_a)$ in (\ref{UK}) that is enforced by a large-$N$ saddle point (Supplementary Information). 
    $G_0$ (black line) is the non-interacting fermion propagator.}
    \label{Dyson}
\end{figure}

\noindent
{\bf Large-$N$ solution.}
The large-$N$ limit of $H$ with interactions defined by (\ref{UK}) and (\ref{Kk}) is described by equations for the Euclidean frequency ($\omega$)/time ($\tau$) 
Green's function $G$ and self energy $\Sigma$ (Fig. \ref{Dyson}):
\bea
&&G^{-1} (k, \omega) = i\omega - \epsilon_k - \Sigma (k, \omega), \label{SYK1} \\
&&\Sigma (k, \tau) = -U^2 \int_{k_1,k_2,k_3} \!\!\!\!\!\!\!\! \delta(\epsilon_k + \epsilon_{k_1} - \epsilon_{k_2} - \epsilon_{k_3}) \nn
&&\times \mathcal{K}_1(k,k_1,k_2,k_3) G(k_1, -\tau) G(k_2, \tau) G(k_3, \tau) \,. \label{SYK2}
\eea
In our solutions of (\ref{SYK1},\ref{SYK2}), we will assume that $\epsilon_k$ is linearized around the Fermi energy $E_F$, so the energy scale $U_{\mathrm{eff}}\equiv U\nu_0^{3/2}\ll E_F, W$, where $\nu_0$ is the density of states at $E_F$. The arguments of $\mathcal{K}_1$ in the above are $k_a = (k, k_1, k_2, k_3)$; in practice, the $\mathcal{K}_1$ factor places upper limits on the values of $|\epsilon_{k_a}-\epsilon_k|$, but the universal structure of the solution for $G(k,\tau)$ described below is not sensitive to its precise form. We will work with 
\bea
\mathcal{K}_1(k,k_1,k_2,k_3) &=& \frac{1}{\Lambda^2}\Xi\left(|\epsilon_{k_1}-\epsilon_k|\right)\Xi\left(|\epsilon_{k_2}-\epsilon_{k_1}|\right)\nn
&\times&\Xi\left(|\epsilon_{k_3}-\epsilon_{k_2}|\right),
\label{intfn}
\eea
where $\Xi(|\epsilon_k|)$ is a smooth function that restricts $|\epsilon_k|\lesssim \Lambda$ and $\Lambda\ll U_{\mathrm{eff}}$. The expression for the self energy becomes
\bea
\Sigma(k,\tau) &=& -\frac{U_{\mathrm{eff}}^2}{\Lambda^2}\int_{-\infty}^{\infty}d\epsilon_{k_2}d\epsilon_{k_3}\Xi\left(|\epsilon_{k_2}-\epsilon_{k_3}|\right)\Xi\left(|\epsilon_{k_3}-\epsilon_k|\right)\nn
&\times&\Xi\left(|\epsilon_{k_2}+\epsilon_{k_3}-2\epsilon_k|\right)G(\epsilon_{k_2},\tau)G(\epsilon_{k_3},\tau)\nn 
&\times& G(\epsilon_{k_2}+\epsilon_{k_3}-\epsilon_k,-\tau). \label{SYK3}
\eea
We also have the particle-hole transmutation relations $G,\Sigma~(-\epsilon_k,\tau)=G,\Sigma~(\epsilon_k,\beta-\tau)$ ($\beta=1/T$).

Our key observation is that the solution of (\ref{SYK1}, \ref{SYK3}) obeys $\omega/T$ scaling for $|\omega|, T, |\epsilon_k|, \Lambda \ll U_{\mathrm{eff}}$, which is the order of scales we shall be mostly interested in in this letter. With the resonant condition applied, this scaling holds all the way down to $T=0$. Moreover, the Green's function has the same form as in the single-site SYK model, but with a momentum dependent dimensionless particle-hole asymmetry parameter $\mathcal{E}_k$; this momentum dependence is sufficient to lead to the remnant Fermi surface (RFS). Specifically, we find under the above specified conditions (Supplementary Information)
\beq
G(k , 0\le \tau < \beta) = \mathcal{A}(\mathcal{E}_k) e^{- 2 \pi \mathcal{E}_k T \tau} \left(\frac{T/U_{\mathrm{eff}}}{\sin(\pi T \tau)} \right)^{2 \Delta}, \label{GPlanckian}
\eeq
and $G(k,\tau+\beta)=-G(k,\tau)$ due to the anticommutation property of fermion operators. The consistency of the $\tau$ dependence in (\ref{GPlanckian}) can be verified by direct substitution into (\ref{SYK1},\ref{SYK2}) using steps similar to those in Ref.~\onlinecite{SS10b}. One finds that the fermion scaling dimension must be $\Delta = 1/4$, as in the in the SYK model with two-body interactions. This value of $\Delta$ is essential to obtain the Planckian behavior of $\tau_\mathrm{tr}$. The exponential $\tau$ dependence in (\ref{GPlanckian}) becomes consistent with (\ref{SYK2}) after imposition of the resonance condition, and a linear dependence of $\mathcal{E}_k$ on the bare quasiparticle energy $\epsilon_k$:
\beq
\mathcal{E}_k = \mathbbm{C} \, \frac{\epsilon_k}{U_{\mathrm{eff}}}\,. \label{fancyC}
\eeq
The dimensionless constant $\mathbbm{C}$ is the not determined by the analytic low energy analysis; its determination requires full numerical solution of (\ref{SYK1},\ref{SYK3}) at all energy scales, and it is given by a slowly varying function of $\Lambda/U_{\mathrm{eff}}$.

In the special case of $\Lambda\rightarrow0$, $\mathcal{K}_1(k,k_1,k_2,k_3)\delta(\epsilon_k+\epsilon_{k_1}-\epsilon_{k_2}-\epsilon_{k_3}) = \delta\left(\epsilon_{k_1}-\epsilon_k\right)\delta\left(\epsilon_{k_2}-\epsilon_{k_1}\right)\delta\left(\epsilon_{k_3}-\epsilon_{k_2}\right)$, and only fermions with the same bare energies interact. For each $\epsilon_k$, (\ref{SYK3}) then reduces to the self-energy of an SYK model with chemical potential $\epsilon_k$ \cite{SY92,SS15}. We then have 
\beq
\mathcal{A}(\mathcal{E}_k)\Bigg|_{\Lambda=0} = -\frac{\pi^{1/4}\cosh^{1/4}(2\pi\mathcal{E}_k)}{\sqrt{1+e^{-4\pi\mathcal{E}_k}}},
\label{AEk}
\eeq
and $\mathbbm{C}\approx 0.41$. For $\Lambda\neq0$, we can only determine $\mathcal{A}(\mathcal{E}_k)$ numerically, but (\ref{AEk}) provides a good approximation (Supplementary Information).
\begin{figure}
    \centering
    \includegraphics[width=0.48\textwidth]{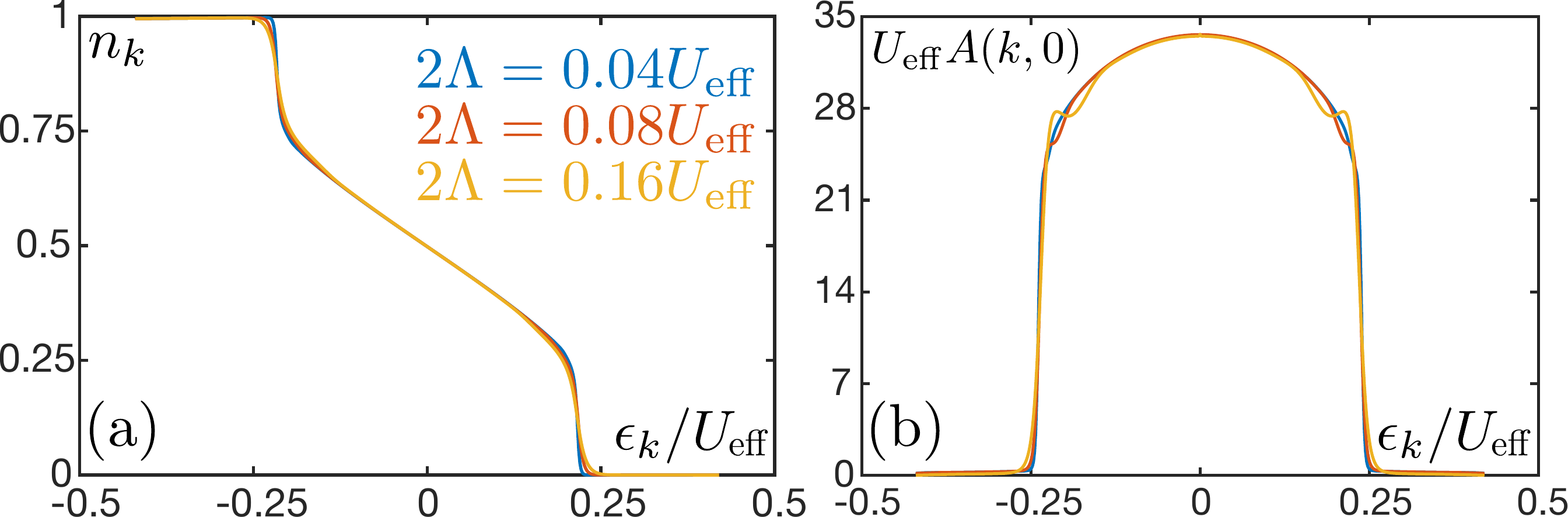}
    \caption{Remnant Fermi surface: (a) The occupancy $n_k$ as a function of $\epsilon_k$, which is exactly $n_k = 1/2$ for $k$ on the RFS ($\epsilon_k=0$). The zero-frequency spectral density  $A(k,0)$ as a function of $\epsilon_k$, which has a maximum for $k$ on the RFS. We use $T/U_\mathrm{eff}=0.01$ and the same color codes for both plots; the function $\Xi(|\epsilon_k|) = 1.866(1-|\epsilon_k|/\Lambda)^2\theta(\Lambda-|\epsilon_k|)$ in (\ref{intfn}), where $\theta$ is the Heaviside step function, with $(1/\Lambda^2)\int_{-\infty}^{\infty}d\epsilon_{k_2}d\epsilon_{k_3}\Xi(|\epsilon_{k_2}-\epsilon_{k_3}|)\Xi(|\epsilon_{k_3}-\epsilon_{k}|)\Xi(|\epsilon_{k_2}+\epsilon_{k_3}-2\epsilon_k|)=1$.}
    \label{remnant}
\end{figure}

\begin{figure}
    \centering
    \includegraphics[width=0.48\textwidth]{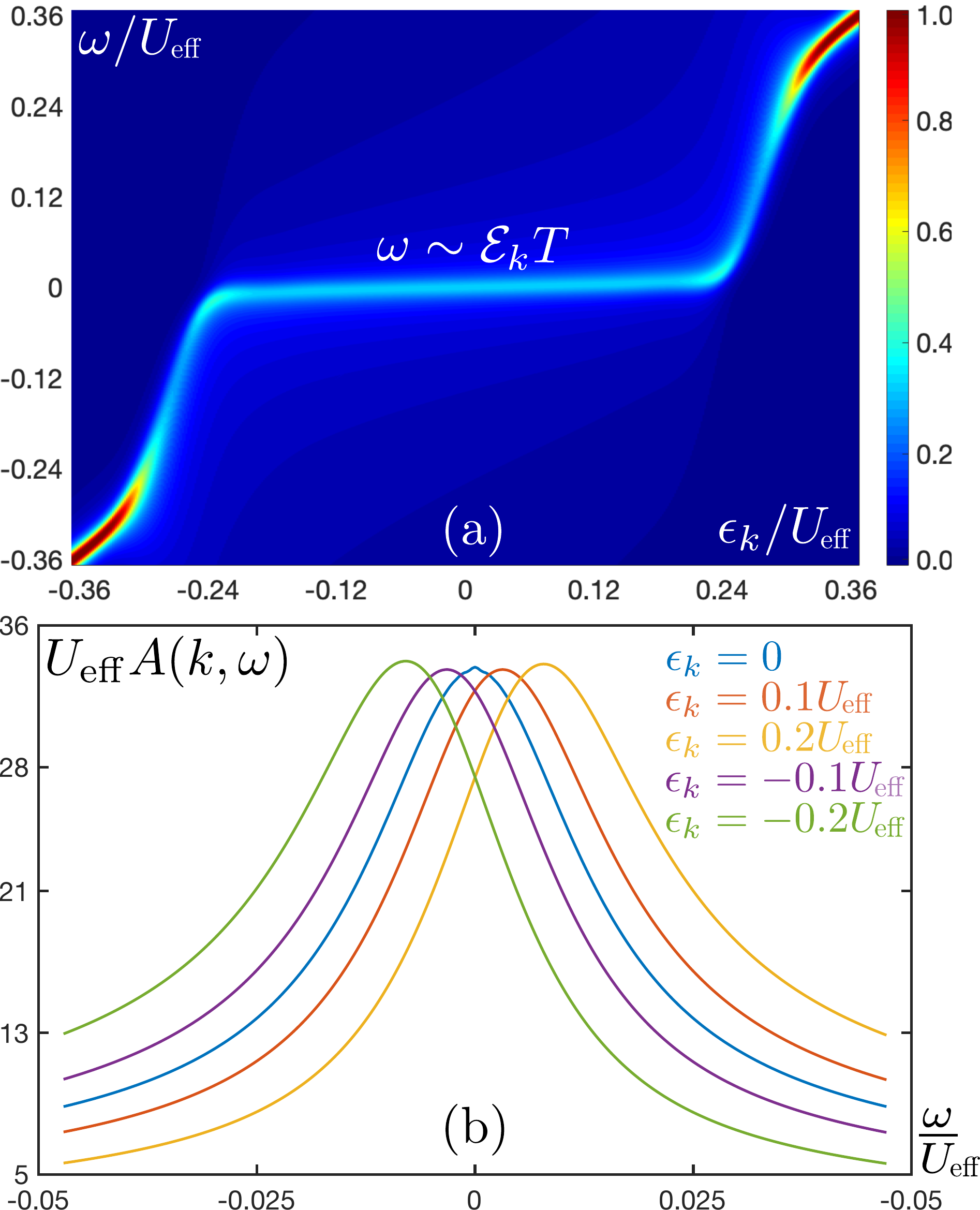}
    \caption{Electron spectral density: (a) The electron spectral density $A(k,\omega)$ (color, normalized scale) as a function of $\epsilon_k$ and frequency $\omega$. Near the RFS the peak disperses as $\omega\sim\mathcal{E}_k T\sim \epsilon_k T/U_\mathrm{eff}$ (see also (b)). Far away from the RFS the peak disperses as $\omega\sim\epsilon_k$ (top right and bottom left corners) and is much sharper. (b) Form of $A(k,\omega)$ near the RFS. We use $T/U_\mathrm{eff}=0.01$ and $2\Lambda=0.16U_\mathrm{eff}$, with the same function $\Xi$ as in Fig. \ref{remnant}.}
    \label{spectral}
\end{figure}

The $k$ dependence of the particle-hole symmetry parameter $\mathcal{E}_k$ in the exponential in (\ref{GPlanckian}) yields an RFS in $G$ (Fig. \ref{remnant}), which is apparent from a computation of the Fourier transform of $G(k,\tau)$ in $\tau$: we find an electron spectral density $A(k,\omega)\equiv-2\mathrm{Im}[G(k,-i\omega+0^+)]$ which is peaked at frequency of order $\mathcal{E}_k T$, and a width of order $T$; see Fig. \ref{spectral}. Using (\ref{fancyC}) as $\epsilon_k$ crosses zero, we therefore find a peak which disperses across the RFS. The volume of the region enclosed by the RFS is not fixed to be that enclosed by the sharp Fermi surface present when $U=0$: The total charge in the system, $q_0$, which is invariant as $U$ is turned on, is given by 
\beq
q_0 = \int_k G(k,\tau=0^-).
\eeq
This is not exactly the volume enclosed by the RFS. However since we have $U_{\mathrm{eff}}\ll E_F$, it is very close to it. The occupancy function $n_k \equiv (1/N)\sum_{\alpha=1}^N\langle c^\dagger_{k,\alpha}c_{k,\alpha}\rangle = G(k,\tau=0^-)$ varies smoothly across the RFS, in contrast to the sharp jump displayed in the non-interacting system. 

The critical solution (\ref{GPlanckian}) does not extend to $\epsilon_k$ arbitrarily far away from the RFS. Consequently, in Fig. \ref{remnant}, $n_k$ reaches $0,1$ for larger $|\epsilon_k|$. This artifact is a consequence of the lack of a non-trivial solution of the single island SYK model for 
$|\epsilon/U| \gtrsim 0.24$, which was noted earlier.
For $|\epsilon_k|\gtrsim\epsilon_k^\ast$, where $\epsilon_k^\ast\sim\mathcal{O}(U_\mathrm{eff})$, (\ref{GPlanckian}) crosses over to a nearly-free solution $G(k,\omega)\sim 1/(i\omega-\epsilon_k)$ instead (Supplementary Information). 
For $\Lambda\rightarrow0$, this crossover becomes a sharp transition \cite{Azeyanagi2018}, whereas for $\Lambda\neq0$ it is a smooth crossover.

It is interesting to compare our Green's function in (\ref{GPlanckian}), with a different higher-dimensional generalization of the SYK solution. In the holographic approach of Refs.~\cite{Faulkner09,Cubrovic:2009ye}, the near-horizon anti-de Sitter AdS$_2$ geometry of a charged black hole in an asymptotically AdS$_D$ ($D \geq 4$) spacetime is identified with that of the complex SYK model \cite{SS10,SS10b}. In this case, the higher-dimensional dispersion yields a near-horizon Green's function whose $\tau$ dependence has the form in (\ref{GPlanckian}), but with a $k$ dependence distinct from ours: their scaling dimension $\Delta$ is dependent upon momentum $k$, while their particle-hole asymmetry $\mathcal{E}$ is momentum independent and determined by the electric field on the surface of the black hole. The placing of the momentum dependence in $\mathcal{E}_k$, and not in $\Delta$, are crucial features of our model, and play a central role in our transport results below, which are very different from the holographic results of Refs.~\cite{Faulkner13}.

\noindent
{\bf Resistivity.} The conductivity $\sigma=1/\rho$ in our model can be computed from the two-point current correlation function using the Kubo formula. In order to do this we note the uniform U(1) current operator, which has two contributions arising from the two lines in (\ref{ressyk});  $I=I_\mathrm{I}+I_\mathrm{II}$, where
\beq
I_\mathrm{I} = e\int_k\sum_{\alpha}\mathbf{v}_k c^\dagger_{k\alpha}c_{k\alpha};~~\mathbf{v}_k = \nabla_k \epsilon_k, 
\label{thecurrent}
\eeq
\bea
&&I_\mathrm{II} = \frac{ie}{(2N)^{3/2}V^{1/2}}\sum_{\alpha_a}\sum_{x_a} U_{\alpha_a}(x_a)c^\dagger_{x_1\alpha_1}c^\dagger_{x_2\alpha_2}c_{x_3\alpha_3}c_{x_4\alpha_4} \nn
&&\times (x_1+x_2-x_3-x_4),
\label{current}
\eea
\beq
U_{\alpha_a}(x_a) = \int_{k_a}U_{\alpha_a}(k_a)e^{i(k_1\cdot x_1+k_2\cdot x_2-k_3\cdot x_3-k_4\cdot x_4)}. \nonumber
\eeq
Here $I_\mathrm{II}$ has been expressed in real ($x_a$) space, and $V$ is the system volume. A consequence of our choice of disorder correlations in (\ref{UK},{\ref{Kk}) is that only $I_\mathrm{I}$ contributes to the current correlation function. Namely, 
\bea
&&\overline{U_{\alpha_a}(x_a)U^\ast_{\alpha'_{a'}}(x'_{a'})} = \mathcal{U}_{\alpha_a \alpha_{a'}'}\Big[\delta_{x_1,x_2}\delta_{x_1,x_3}\nn
&&\times\delta_{x_1,x_4}\mathcal{F}(x_1,x'_{a'}) + (x_a\leftrightarrow x'_{a'})\Big],
\label{UX1}
\eea
with
\bea
&&\mathcal{F}(x_1,x'_{a'}) = \frac{1}{2}\int_{k'_{a'}}\mathcal{K}_1(k'_{a'})\delta(\epsilon_{k'_1}+\epsilon_{k'_2}-\epsilon_{k'_3}-\epsilon_{k'_4})\nn
&&\times e^{i(k'_1\cdot(x_1-x'_1)+k'_2\cdot(x_1-x'_2)-k'_3\cdot(x_1-x'_3)-k'_4\cdot(x_1-x'_4))},
\label{UX2}
\eea
and $\mathcal{F}^\ast(x_1,x'_{a'})=\mathcal{F}(x_1,x'_{a'})$. Therefore, under the disorder average, and in the large-$N$ limit, the two-point function of $I_\mathrm{II}$ vanishes as either all $x_a$ or all $x'_{a'}$ have to be the same. There is also no two-point cross correlation of $I_\mathrm{I}$ and $I_\mathrm{II}$ in the large-$N$ limit, so we only need to compute the two-point function of $I_\mathrm{I}$. 

In the large-$N$ limit, the two point function of $I_\mathrm{I}$ is given by the bubble diagram (Fig. \ref{siggraph}) plus a series of ladder diagrams. The ladder diagrams however vanish due to inversion symmetry. We thus have 
\bea
&&\sigma = \lim_{\Omega\rightarrow0}\frac{\mathrm{Im}[\langle I_\mathrm{I}\cdot I_\mathrm{I}\rangle^R(\Omega)]}{\Omega d} = \int_{-\infty}^{\infty}d\epsilon\int_k \frac{|\nabla_k\epsilon_k|^2}{d}\delta(\epsilon_k-\epsilon) \nn
&&\times\frac{Ne^2}{T}\int_{-\infty}^{\infty}\frac{d\omega}{16\pi}\mathrm{sech}^2\left(\frac{\omega}{2T}\right)A^2(\epsilon,\omega).
\label{sigma1}
\eea
If we linearize $\epsilon_k$ around the RFS, we can replace 
\beq
\int_k \frac{|\nabla_k\epsilon_k|^2}{d}\delta(\epsilon_k-\epsilon) = \oint_\mathrm{RFS} \frac{d^{d-1}a_k}{(2\pi)^d}\frac{|\nabla_k\epsilon_k|}{d} \equiv \frac{V_{RFS}}{m^\ast},
\label{msta}
\eeq
where $d^{d-1}a_k$ is the local area element on the RFS and $V_\mathrm{RFS}=\int_k \theta(-\epsilon_k)$ is the volume of the RFS. As noted earlier, since $U_\mathrm{eff}\ll E_F$, the carrier density is $n = NV_\mathrm{RFS}$. The second equality in (\ref{msta}) provides our definition of $m^\ast$, which coincides with the traditional definition for $\epsilon_k = k^2/(2m^\ast)-E_F$ in any number of dimensions $d$. 
\begin{figure}
    \centering
    \includegraphics[width=0.4\textwidth]{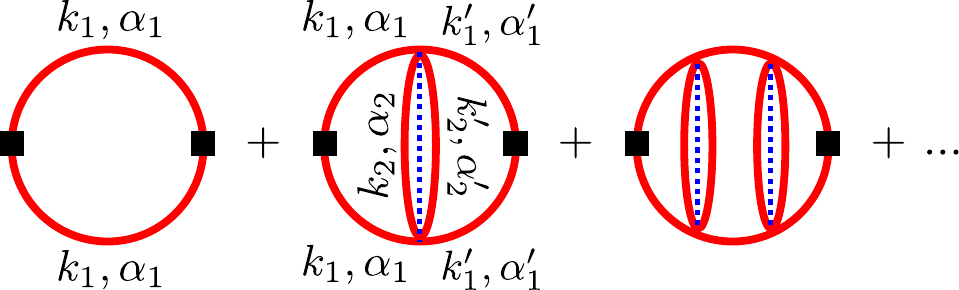}
    \caption{Contributions to the current correlation function in the large-$N$ limit. The boxes denote $I_\mathrm{I}$ vertices, and the lines are defined as in Fig. \ref{Dyson}. The ladder diagrams vanish because the directions of the momenta $k_1,k_1'$ flowing through the current vertices are not correlated with each other and $v_{-k}=-v_{k}$, $\epsilon_{-k}=\epsilon_k$.}
    \label{siggraph}
\end{figure}
The spectral function obtained from (\ref{GPlanckian}) has the low-energy form
\bea
&&A(\epsilon_k,\omega) = -2\mathcal{A}(\mathcal{E}_k) \nn
&&\times\mathrm{Im}\left[\frac{e^{\frac{i \pi }{4}} \left(1+i e^{-2\pi\mathcal{E}_k}\right) \Gamma \left(\frac{i(2\pi\mathcal{E}_kT-\omega)}{2\pi  T}+\frac{\pi}{4}\right)}{\Gamma \left(\frac{i(2\pi\mathcal{E}_kT-\omega)}{2\pi T}+\frac{3}{4}\right)\sqrt{2\pi TU_\mathrm{eff}}}\right], \label{AA}
\eea

We now describe a key feature of (\ref{sigma1}): the insensitivity of the conductivity to the value of $U_{\mathrm{eff}}$.
Notice that the expression for the conductivity in (\ref{sigma1}) has two explicit dependencies on $U_{\mathrm{eff}}$: from the spectral density in (\ref{AA}); and from that in $\mathcal{E}_k$ which is a linear function of $\epsilon_k$ in (\ref{fancyC}),
but is more generally given by $\mathcal{E}_k=g(\epsilon_k/U_\mathrm{eff})$ for some function $g$.
Therefore, the explicit dependence on $U_\mathrm{eff}$ disappears from (\ref{sigma1}) by scaling the integrand $\epsilon\rightarrow \epsilon U_\mathrm{eff}$. There is some implicit dependence on the ratio $\Lambda/U_{\mathrm{eff}}$ arising from the functional forms of $\mathcal{A}$ and $g$, but our numerical computations show that this dependence is negligible, and not more than a few percent. 

A more serious non-universality of (\ref{sigma1}) is that arising from the artifact of the single island SYK model: the crossover to nearly-free fermions with $n_k = 0,1$ for $|\epsilon_k| \gtrsim\epsilon^\ast_k$. We do not expect this feature to be present in more realistic microscopic models, so a reasonable strategy is to work around this feature by assuming that the linearized expression for $\mathcal{E}_k$ in (\ref{fancyC}) holds for the full range of $\epsilon$ integration in (\ref{sigma1}); in other words, we are focusing on the physics proximate to the RFS, and ignoring unphysical features of our simple model far from the RFS. We then find that the resulting resistivity $\rho=1/\sigma$ has precisely the Drude form in (\ref{drude}), and the transport time is given by the Planckian form in (\ref{planck}). The numerical constant $f$ is a very weak function of $\Lambda/U_{\mathrm{eff}}$, and takes the value $f\approx 1.11$
in the limit $\Lambda\rightarrow0$ of independent momentum shells. If we include the artifact of the crossover far from the RFS, then the Planckian forms in (\ref{drude}) and (\ref{planck}) continue to hold, and our numerical computations yield $f \approx 4.98$, and again fairly insensitive to $\Lambda/U_{\mathrm{eff}}$.

We also computed $\sigma$ by directly inserting the numerically determined spectral density into (\ref{sigma1},\ref{msta}), without restricting the integrals (Fig. \ref{rhoT}). For $T\ll U_\mathrm{eff}$ we find that the above transport results work well.
\begin{figure}
    \centering
    \includegraphics[width=0.48\textwidth]{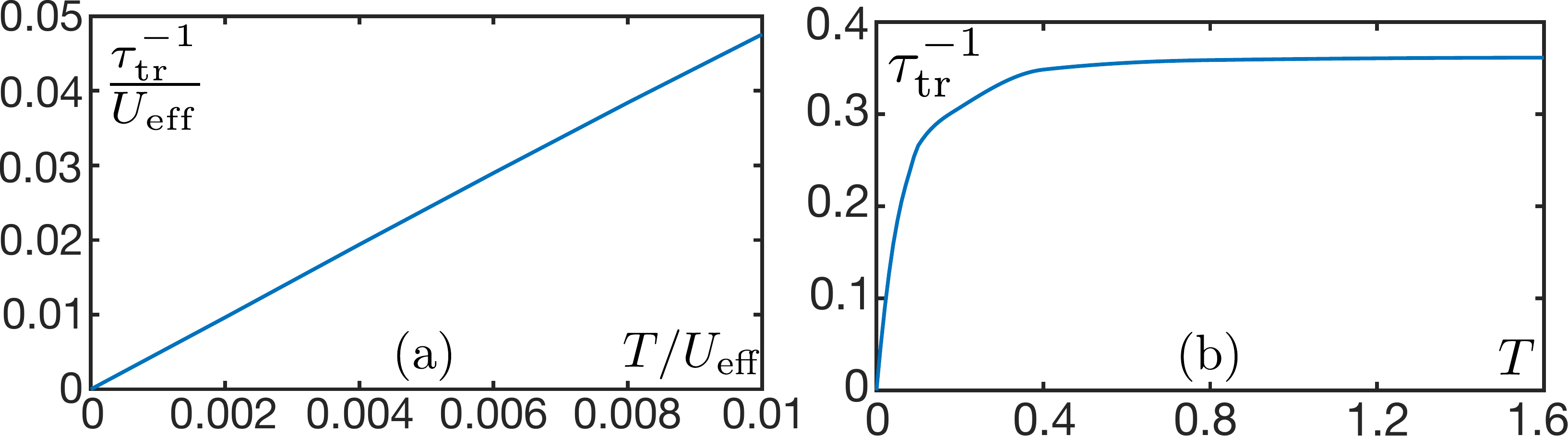}
    \caption{Numerically calculated transport scattering rate $\tau_\mathrm{tr}^{-1}$ vs. temperature $T$. (a) For $T\ll U_\mathrm{eff}$, we find the Planckian dependence. We have $2\Lambda=0.04U_\mathrm{eff}$, and find $f\approx 4.8$. (b) As $T$ approaches and exceeds $U_\mathrm{eff}$, $\tau_\mathrm{tr}^{-1}$ saturates to $\tau_\mathrm{tr}^{-1}\sim \mathcal{O}(U_\mathrm{eff})$ (plotted for $U_\mathrm{eff}=1$): In this high-temperature regime, $A(\epsilon,\omega)$ has a peak in $\omega$ of width $\sim U_\mathrm{eff}$ and height $\sim U_\mathrm{eff}^{-1}$, which is centered at $\omega\ll T$ for $\epsilon\ll T$. These properties then produce the observed behavior from (\ref{sigma1}), in which the $\epsilon$ integral is now effectively cut off by $T$ instead of $U_\mathrm{eff}$. Note that $\hbar=k_B=1$ in these plots.}
    \label{rhoT}
\end{figure}

\noindent
{\bf Outlook.} We have introduced resonant SYK models: these have the unique features of being solvable while realizing compressible states of quantum matter, in non-zero spatial dimensions, whose spectral functions obey $\hbar\omega/(k_B T)$ scaling  as $T \rightarrow 0$. We presented a renormalization group motivation for the resonance condition, but within the large-$N$ limit, the resonance appears as a fine-tuning condition designed to retain $\hbar\omega/(k_B T)$ scaling down to $T = 0$ even in the presence of a dispersion. However, then many other desirable features follow: we obtain a Planckian metal with a large RFS at $\epsilon_k = 0$, and an effective mass $m^\ast$ defined by the dispersion of $\epsilon_k$, with a resistivity $\rho \sim (m^\ast/(ne^2)) k_B T/\hbar$ {\it independent\/} of the strength of interactions.

It would be interesting to extend our model to include disorder in single-particle terms in the Hamiltonian. For the resonant SYK interactions described above, we expect a crossover to a $T$-independent residual resistivity at low enough $T$. Refs.~\cite{Trivedi19,Trivedi19a} examine models with single-particle disorder and local interactions by numerically summing the `melonic' diagrams of Fig. \ref{Dyson} without averaging over disorder, and without a resonance condition: they find that Planckian transport exists over a wide range of $T$.

Photoemission experiments can test features of the spectral function in Fig. \ref{spectral} of the Planckian metal: an energy width $\sim T$ near the RFS, and a `kink' in the dispersion close to the RFS with the apparent Fermi velocity becoming proportional to $T$.

\noindent
{\bf Acknowledgements.} We thank Ganapathy Baskaran, Kyungmin Lee, Juan Maldacena, and Nandini Trivedi for inspirational discussions. This research was supported by the US Department of Energy under Grant No. DE-SC0019030.

\bibliography{syk}

\begin{thebibliography}{31}%
\makeatletter
\providecommand \@ifxundefined [1]{%
 \@ifx{#1\undefined}
}%
\providecommand \@ifnum [1]{%
 \ifnum #1\expandafter \@firstoftwo
 \else \expandafter \@secondoftwo
 \fi
}%
\providecommand \@ifx [1]{%
 \ifx #1\expandafter \@firstoftwo
 \else \expandafter \@secondoftwo
 \fi
}%
\providecommand \natexlab [1]{#1}%
\providecommand \enquote  [1]{``#1''}%
\providecommand \bibnamefont  [1]{#1}%
\providecommand \bibfnamefont [1]{#1}%
\providecommand \citenamefont [1]{#1}%
\providecommand \href@noop [0]{\@secondoftwo}%
\providecommand \href [0]{\begingroup \@sanitize@url \@href}%
\providecommand \@href[1]{\@@startlink{#1}\@@href}%
\providecommand \@@href[1]{\endgroup#1\@@endlink}%
\providecommand \@sanitize@url [0]{\catcode `\\12\catcode `\$12\catcode
  `\&12\catcode `\#12\catcode `\^12\catcode `\_12\catcode `\%12\relax}%
\providecommand \@@startlink[1]{}%
\providecommand \@@endlink[0]{}%
\providecommand \url  [0]{\begingroup\@sanitize@url \@url }%
\providecommand \@url [1]{\endgroup\@href {#1}{\urlprefix }}%
\providecommand \urlprefix  [0]{URL }%
\providecommand \Eprint [0]{\href }%
\providecommand \doibase [0]{http://dx.doi.org/}%
\providecommand \selectlanguage [0]{\@gobble}%
\providecommand \bibinfo  [0]{\@secondoftwo}%
\providecommand \bibfield  [0]{\@secondoftwo}%
\providecommand \translation [1]{[#1]}%
\providecommand \BibitemOpen [0]{}%
\providecommand \bibitemStop [0]{}%
\providecommand \bibitemNoStop [0]{.\EOS\space}%
\providecommand \EOS [0]{\spacefactor3000\relax}%
\providecommand \BibitemShut  [1]{\csname bibitem#1\endcsname}%
\let\auto@bib@innerbib\@empty
\bibitem [{\citenamefont {Takagi}\ \emph {et~al.}(1992)\citenamefont {Takagi},
  \citenamefont {Batlogg}, \citenamefont {Kao}, \citenamefont {Kwo},
  \citenamefont {Cava}, \citenamefont {Krajewski},\ and\ \citenamefont
  {Peck}}]{Takagi92}%
  \BibitemOpen
  \bibfield  {author} {\bibinfo {author} {\bibfnamefont {H.}~\bibnamefont
  {Takagi}}, \bibinfo {author} {\bibfnamefont {B.}~\bibnamefont {Batlogg}},
  \bibinfo {author} {\bibfnamefont {H.~L.}\ \bibnamefont {Kao}}, \bibinfo
  {author} {\bibfnamefont {J.}~\bibnamefont {Kwo}}, \bibinfo {author}
  {\bibfnamefont {R.~J.}\ \bibnamefont {Cava}}, \bibinfo {author}
  {\bibfnamefont {J.~J.}\ \bibnamefont {Krajewski}}, \ and\ \bibinfo {author}
  {\bibfnamefont {W.~F.}\ \bibnamefont {Peck}},\ }\bibfield  {title} {\enquote
  {\bibinfo {title} {{Systematic evolution of temperature-dependent resistivity
  in
  ${\mathrm{La}}_{2\mathrm{\ensuremath{-}}\mathit{x}}$${\mathrm{Sr}}_{\mathit{x}}$${\mathrm{CuO}}_{4}$}},}\
  }\href {\doibase 10.1103/PhysRevLett.69.2975} {\bibfield  {journal} {\bibinfo
   {journal} {Phys. Rev. Lett.}\ }\textbf {\bibinfo {volume} {69}},\ \bibinfo
  {pages} {2975} (\bibinfo {year} {1992})}\BibitemShut {NoStop}%
\bibitem [{\citenamefont {{Taillefer}}(2010)}]{Taillefer10}%
  \BibitemOpen
  \bibfield  {author} {\bibinfo {author} {\bibfnamefont {L.}~\bibnamefont
  {{Taillefer}}},\ }\bibfield  {title} {\enquote {\bibinfo {title} {{Scattering
  and Pairing in Cuprate Superconductors}},}\ }\href {\doibase
  10.1146/annurev-conmatphys-070909-104117} {\bibfield  {journal} {\bibinfo
  {journal} {Annual Review of Condensed Matter Physics}\ }\textbf {\bibinfo
  {volume} {1}},\ \bibinfo {pages} {51} (\bibinfo {year} {2010})},\ \Eprint
  {http://arxiv.org/abs/1003.2972} {arXiv:1003.2972 [cond-mat.supr-con]}
  \BibitemShut {NoStop}%
\bibitem [{\citenamefont {Bruin}\ \emph {et~al.}(2013)\citenamefont {Bruin},
  \citenamefont {Sakai}, \citenamefont {Perry},\ and\ \citenamefont
  {Mackenzie}}]{MacKenzie13}%
  \BibitemOpen
  \bibfield  {author} {\bibinfo {author} {\bibfnamefont {J.~A.~N.}\
  \bibnamefont {Bruin}}, \bibinfo {author} {\bibfnamefont {H.}~\bibnamefont
  {Sakai}}, \bibinfo {author} {\bibfnamefont {R.~S.}\ \bibnamefont {Perry}}, \
  and\ \bibinfo {author} {\bibfnamefont {A.~P.}\ \bibnamefont {Mackenzie}},\
  }\bibfield  {title} {\enquote {\bibinfo {title} {{Similarity of Scattering
  Rates in Metals Showing $T$-Linear Resistivity}},}\ }\href {\doibase
  10.1126/science.1227612} {\bibfield  {journal} {\bibinfo  {journal}
  {Science}\ }\textbf {\bibinfo {volume} {339}},\ \bibinfo {pages} {804}
  (\bibinfo {year} {2013})}\BibitemShut {NoStop}%
\bibitem [{\citenamefont {{Legros}}\ \emph {et~al.}(2018)\citenamefont
  {{Legros}}, \citenamefont {{Benhabib}}, \citenamefont {{Tabis}},
  \citenamefont {{Lalibert{\'e}}}, \citenamefont {{Dion}}, \citenamefont
  {{Lizaire}}, \citenamefont {{Vignolle}}, \citenamefont {{Vignolles}},
  \citenamefont {{Raffy}}, \citenamefont {{Li}}, \citenamefont
  {{Auban-Senzier}}, \citenamefont {{Doiron-Leyraud}}, \citenamefont
  {{Fournier}}, \citenamefont {{Colson}}, \citenamefont {{Taillefer}},\ and\
  \citenamefont {{Proust}}}]{Legros18}%
  \BibitemOpen
  \bibfield  {author} {\bibinfo {author} {\bibfnamefont {A.}~\bibnamefont
  {{Legros}}}, \bibinfo {author} {\bibfnamefont {S.}~\bibnamefont
  {{Benhabib}}}, \bibinfo {author} {\bibfnamefont {W.}~\bibnamefont {{Tabis}}},
  \bibinfo {author} {\bibfnamefont {F.}~\bibnamefont {{Lalibert{\'e}}}},
  \bibinfo {author} {\bibfnamefont {M.}~\bibnamefont {{Dion}}}, \bibinfo
  {author} {\bibfnamefont {M.}~\bibnamefont {{Lizaire}}}, \bibinfo {author}
  {\bibfnamefont {B.}~\bibnamefont {{Vignolle}}}, \bibinfo {author}
  {\bibfnamefont {D.}~\bibnamefont {{Vignolles}}}, \bibinfo {author}
  {\bibfnamefont {H.}~\bibnamefont {{Raffy}}}, \bibinfo {author} {\bibfnamefont
  {Z.~Z.}\ \bibnamefont {{Li}}}, \bibinfo {author} {\bibfnamefont
  {P.}~\bibnamefont {{Auban-Senzier}}}, \bibinfo {author} {\bibfnamefont
  {N.}~\bibnamefont {{Doiron-Leyraud}}}, \bibinfo {author} {\bibfnamefont
  {P.}~\bibnamefont {{Fournier}}}, \bibinfo {author} {\bibfnamefont
  {D.}~\bibnamefont {{Colson}}}, \bibinfo {author} {\bibfnamefont
  {L.}~\bibnamefont {{Taillefer}}}, \ and\ \bibinfo {author} {\bibfnamefont
  {C.}~\bibnamefont {{Proust}}},\ }\bibfield  {title} {\enquote {\bibinfo
  {title} {{Universal $T$-linear resistivity and Planckian dissipation in
  overdoped cuprates}},}\ }\href {\doibase 10.1038/s41567-018-0334-2}
  {\bibfield  {journal} {\bibinfo  {journal} {Nature Physics}\ }\textbf
  {\bibinfo {volume} {15}},\ \bibinfo {pages} {142} (\bibinfo {year} {2018})},\
  \Eprint {http://arxiv.org/abs/1805.02512} {arXiv:1805.02512
  [cond-mat.supr-con]} \BibitemShut {NoStop}%
\bibitem [{\citenamefont {{Nakajima}}\ \emph {et~al.}(2019)\citenamefont
  {{Nakajima}}, \citenamefont {{Metz}}, \citenamefont {{Eckberg}},
  \citenamefont {{Kirshenbaum}}, \citenamefont {{Hughes}}, \citenamefont
  {{Wang}}, \citenamefont {{Wang}}, \citenamefont {{Saha}}, \citenamefont
  {{Liu}}, \citenamefont {{Butch}}, \citenamefont {{Campbell}}, \citenamefont
  {{Eo}}, \citenamefont {{Graf}}, \citenamefont {{Liu}}, \citenamefont
  {{Borisenko}}, \citenamefont {{Zavalij}},\ and\ \citenamefont
  {{Paglione}}}]{Paglione19}%
  \BibitemOpen
  \bibfield  {author} {\bibinfo {author} {\bibfnamefont {Y.}~\bibnamefont
  {{Nakajima}}}, \bibinfo {author} {\bibfnamefont {T.}~\bibnamefont {{Metz}}},
  \bibinfo {author} {\bibfnamefont {C.}~\bibnamefont {{Eckberg}}}, \bibinfo
  {author} {\bibfnamefont {K.}~\bibnamefont {{Kirshenbaum}}}, \bibinfo {author}
  {\bibfnamefont {A.}~\bibnamefont {{Hughes}}}, \bibinfo {author}
  {\bibfnamefont {R.}~\bibnamefont {{Wang}}}, \bibinfo {author} {\bibfnamefont
  {L.}~\bibnamefont {{Wang}}}, \bibinfo {author} {\bibfnamefont {S.~R.}\
  \bibnamefont {{Saha}}}, \bibinfo {author} {\bibfnamefont {I.-L.}\
  \bibnamefont {{Liu}}}, \bibinfo {author} {\bibfnamefont {N.~P.}\ \bibnamefont
  {{Butch}}}, \bibinfo {author} {\bibfnamefont {D.}~\bibnamefont {{Campbell}}},
  \bibinfo {author} {\bibfnamefont {Y.~S.}\ \bibnamefont {{Eo}}}, \bibinfo
  {author} {\bibfnamefont {D.}~\bibnamefont {{Graf}}}, \bibinfo {author}
  {\bibfnamefont {Z.}~\bibnamefont {{Liu}}}, \bibinfo {author} {\bibfnamefont
  {S.~V.}\ \bibnamefont {{Borisenko}}}, \bibinfo {author} {\bibfnamefont
  {P.~Y.}\ \bibnamefont {{Zavalij}}}, \ and\ \bibinfo {author} {\bibfnamefont
  {J.}~\bibnamefont {{Paglione}}},\ }\bibfield  {title} {\enquote {\bibinfo
  {title} {{Planckian dissipation and scale invariance in a quantum-critical
  disordered pnictide}},}\ }\href@noop {} {\bibfield  {journal} {\bibinfo
  {journal} {arXiv e-prints}\ } (\bibinfo {year} {2019})},\ \Eprint
  {http://arxiv.org/abs/1902.01034} {arXiv:1902.01034 [cond-mat.str-el]}
  \BibitemShut {NoStop}%
\bibitem [{\citenamefont {{Cao}}\ \emph {et~al.}(2019)\citenamefont {{Cao}},
  \citenamefont {{Chowdhury}}, \citenamefont {{Rodan-Legrain}}, \citenamefont
  {{Rubies-Bigord{\`a}}}, \citenamefont {{Watanabe}}, \citenamefont
  {{Taniguchi}}, \citenamefont {{Senthil}},\ and\ \citenamefont
  {{Jarillo-Herrero}}}]{Pablo19}%
  \BibitemOpen
  \bibfield  {author} {\bibinfo {author} {\bibfnamefont {Y.}~\bibnamefont
  {{Cao}}}, \bibinfo {author} {\bibfnamefont {D.}~\bibnamefont {{Chowdhury}}},
  \bibinfo {author} {\bibfnamefont {D.}~\bibnamefont {{Rodan-Legrain}}},
  \bibinfo {author} {\bibfnamefont {O.}~\bibnamefont {{Rubies-Bigord{\`a}}}},
  \bibinfo {author} {\bibfnamefont {K.}~\bibnamefont {{Watanabe}}}, \bibinfo
  {author} {\bibfnamefont {T.}~\bibnamefont {{Taniguchi}}}, \bibinfo {author}
  {\bibfnamefont {T.}~\bibnamefont {{Senthil}}}, \ and\ \bibinfo {author}
  {\bibfnamefont {P.}~\bibnamefont {{Jarillo-Herrero}}},\ }\bibfield  {title}
  {\enquote {\bibinfo {title} {{Strange metal in magic-angle graphene with near
  Planckian dissipation}},}\ }\href@noop {} {\bibfield  {journal} {\bibinfo
  {journal} {arXiv e-prints}\ } (\bibinfo {year} {2019})},\ \Eprint
  {http://arxiv.org/abs/1901.03710} {arXiv:1901.03710 [cond-mat.str-el]}
  \BibitemShut {NoStop}%
\bibitem [{\citenamefont {{Brown}}\ \emph {et~al.}(2019)\citenamefont
  {{Brown}}, \citenamefont {{Mitra}}, \citenamefont {{Guardado-Sanchez}},
  \citenamefont {{Nourafkan}}, \citenamefont {{Reymbaut}}, \citenamefont
  {{H{\'e}bert}}, \citenamefont {{Bergeron}}, \citenamefont {{Tremblay}},
  \citenamefont {{Kokalj}}, \citenamefont {{Huse}}, \citenamefont
  {{Schau{\ss}}},\ and\ \citenamefont {{Bakr}}}]{Bakr19}%
  \BibitemOpen
  \bibfield  {author} {\bibinfo {author} {\bibfnamefont {P.~T.}\ \bibnamefont
  {{Brown}}}, \bibinfo {author} {\bibfnamefont {D.}~\bibnamefont {{Mitra}}},
  \bibinfo {author} {\bibfnamefont {E.}~\bibnamefont {{Guardado-Sanchez}}},
  \bibinfo {author} {\bibfnamefont {R.}~\bibnamefont {{Nourafkan}}}, \bibinfo
  {author} {\bibfnamefont {A.}~\bibnamefont {{Reymbaut}}}, \bibinfo {author}
  {\bibfnamefont {C.-D.}\ \bibnamefont {{H{\'e}bert}}}, \bibinfo {author}
  {\bibfnamefont {S.}~\bibnamefont {{Bergeron}}}, \bibinfo {author}
  {\bibfnamefont {A.~M.~S.}\ \bibnamefont {{Tremblay}}}, \bibinfo {author}
  {\bibfnamefont {J.}~\bibnamefont {{Kokalj}}}, \bibinfo {author}
  {\bibfnamefont {D.~A.}\ \bibnamefont {{Huse}}}, \bibinfo {author}
  {\bibfnamefont {P.}~\bibnamefont {{Schau{\ss}}}}, \ and\ \bibinfo {author}
  {\bibfnamefont {W.~S.}\ \bibnamefont {{Bakr}}},\ }\bibfield  {title}
  {\enquote {\bibinfo {title} {{Bad metallic transport in a cold atom
  Fermi-Hubbard system}},}\ }\href {\doibase 10.1126/science.aat4134}
  {\bibfield  {journal} {\bibinfo  {journal} {Science}\ }\textbf {\bibinfo
  {volume} {363}},\ \bibinfo {pages} {379} (\bibinfo {year} {2019})},\ \Eprint
  {http://arxiv.org/abs/1802.09456} {arXiv:1802.09456 [cond-mat.quant-gas]}
  \BibitemShut {NoStop}%
\bibitem [{\citenamefont {{Parcollet}}\ and\ \citenamefont
  {{Georges}}(1999)}]{PG98}%
  \BibitemOpen
  \bibfield  {author} {\bibinfo {author} {\bibfnamefont {O.}~\bibnamefont
  {{Parcollet}}}\ and\ \bibinfo {author} {\bibfnamefont {A.}~\bibnamefont
  {{Georges}}},\ }\bibfield  {title} {\enquote {\bibinfo {title}
  {{Non-Fermi-liquid regime of a doped Mott insulator}},}\ }\href {\doibase
  10.1103/PhysRevB.59.5341} {\bibfield  {journal} {\bibinfo  {journal} {Phys.
  Rev. B}\ }\textbf {\bibinfo {volume} {59}},\ \bibinfo {pages} {5341}
  (\bibinfo {year} {1999})},\ \Eprint {http://arxiv.org/abs/cond-mat/9806119}
  {cond-mat/9806119} \BibitemShut {NoStop}%
\bibitem [{\citenamefont {{Zhang}}(2017)}]{Zhang2017}%
  \BibitemOpen
  \bibfield  {author} {\bibinfo {author} {\bibfnamefont {P.}~\bibnamefont
  {{Zhang}}},\ }\bibfield  {title} {\enquote {\bibinfo {title} {{Dispersive
  Sachdev-Ye-Kitaev model: Band structure and quantum chaos}},}\ }\href
  {\doibase 10.1103/PhysRevB.96.205138} {\bibfield  {journal} {\bibinfo
  {journal} {Phys. Rev. B}\ }\textbf {\bibinfo {volume} {96}},\ \bibinfo {eid}
  {205138} (\bibinfo {year} {2017})},\ \Eprint
  {http://arxiv.org/abs/1707.09589} {arXiv:1707.09589 [cond-mat.str-el]}
  \BibitemShut {NoStop}%
\bibitem [{\citenamefont {{Song}}\ \emph {et~al.}(2017)\citenamefont {{Song}},
  \citenamefont {{Jian}},\ and\ \citenamefont {{Balents}}}]{Balents2017}%
  \BibitemOpen
  \bibfield  {author} {\bibinfo {author} {\bibfnamefont {X.-Y.}\ \bibnamefont
  {{Song}}}, \bibinfo {author} {\bibfnamefont {C.-M.}\ \bibnamefont {{Jian}}},
  \ and\ \bibinfo {author} {\bibfnamefont {L.}~\bibnamefont {{Balents}}},\
  }\bibfield  {title} {\enquote {\bibinfo {title} {{Strongly Correlated Metal
  Built from Sachdev-Ye-Kitaev Models}},}\ }\href {\doibase
  10.1103/PhysRevLett.119.216601} {\bibfield  {journal} {\bibinfo  {journal}
  {Phys. Rev. Lett.}\ }\textbf {\bibinfo {volume} {119}},\ \bibinfo {eid}
  {216601} (\bibinfo {year} {2017})},\ \Eprint
  {http://arxiv.org/abs/1705.00117} {arXiv:1705.00117 [cond-mat.str-el]}
  \BibitemShut {NoStop}%
\bibitem [{\citenamefont {Patel}\ \emph
  {et~al.}(2018{\natexlab{a}})\citenamefont {Patel}, \citenamefont {McGreevy},
  \citenamefont {Arovas},\ and\ \citenamefont {Sachdev}}]{Patel2017}%
  \BibitemOpen
  \bibfield  {author} {\bibinfo {author} {\bibfnamefont {A.~A.}\ \bibnamefont
  {Patel}}, \bibinfo {author} {\bibfnamefont {J.}~\bibnamefont {McGreevy}},
  \bibinfo {author} {\bibfnamefont {D.~P.}\ \bibnamefont {Arovas}}, \ and\
  \bibinfo {author} {\bibfnamefont {S.}~\bibnamefont {Sachdev}},\ }\bibfield
  {title} {\enquote {\bibinfo {title} {{Magnetotransport in a model of a
  disordered strange metal}},}\ }\href {\doibase 10.1103/PhysRevX.8.021049}
  {\bibfield  {journal} {\bibinfo  {journal} {Phys. Rev. X}\ }\textbf {\bibinfo
  {volume} {8}},\ \bibinfo {pages} {021049} (\bibinfo {year}
  {2018}{\natexlab{a}})},\ \Eprint {http://arxiv.org/abs/1712.05026}
  {arXiv:1712.05026 [cond-mat.str-el]} \BibitemShut {NoStop}%
\bibitem [{\citenamefont {Chowdhury}\ \emph {et~al.}(2018)\citenamefont
  {Chowdhury}, \citenamefont {Werman}, \citenamefont {Berg},\ and\
  \citenamefont {Senthil}}]{Chowdhury2018}%
  \BibitemOpen
  \bibfield  {author} {\bibinfo {author} {\bibfnamefont {D.}~\bibnamefont
  {Chowdhury}}, \bibinfo {author} {\bibfnamefont {Y.}~\bibnamefont {Werman}},
  \bibinfo {author} {\bibfnamefont {E.}~\bibnamefont {Berg}}, \ and\ \bibinfo
  {author} {\bibfnamefont {T.}~\bibnamefont {Senthil}},\ }\bibfield  {title}
  {\enquote {\bibinfo {title} {{Translationally invariant non-Fermi liquid
  metals with critical Fermi-surfaces: Solvable models}},}\ }\href {\doibase
  10.1103/PhysRevX.8.031024} {\bibfield  {journal} {\bibinfo  {journal} {Phys.
  Rev. X}\ }\textbf {\bibinfo {volume} {8}},\ \bibinfo {pages} {031024}
  (\bibinfo {year} {2018})},\ \Eprint {http://arxiv.org/abs/1801.06178}
  {arXiv:1801.06178 [cond-mat.str-el]} \BibitemShut {NoStop}%
\bibitem [{\citenamefont {Patel}\ \emph
  {et~al.}(2018{\natexlab{b}})\citenamefont {Patel}, \citenamefont {Lawler},\
  and\ \citenamefont {Kim}}]{PatelKim2018}%
  \BibitemOpen
  \bibfield  {author} {\bibinfo {author} {\bibfnamefont {A.~A.}\ \bibnamefont
  {Patel}}, \bibinfo {author} {\bibfnamefont {M.~J.}\ \bibnamefont {Lawler}}, \
  and\ \bibinfo {author} {\bibfnamefont {E.-A.}\ \bibnamefont {Kim}},\
  }\bibfield  {title} {\enquote {\bibinfo {title} {Coherent superconductivity
  with a large gap ratio from incoherent metals},}\ }\href {\doibase
  10.1103/PhysRevLett.121.187001} {\bibfield  {journal} {\bibinfo  {journal}
  {Phys. Rev. Lett.}\ }\textbf {\bibinfo {volume} {121}},\ \bibinfo {pages}
  {187001} (\bibinfo {year} {2018}{\natexlab{b}})}\BibitemShut {NoStop}%
\bibitem [{\citenamefont {{Sachdev}}\ and\ \citenamefont {{Ye}}(1993)}]{SY92}%
  \BibitemOpen
  \bibfield  {author} {\bibinfo {author} {\bibfnamefont {S.}~\bibnamefont
  {{Sachdev}}}\ and\ \bibinfo {author} {\bibfnamefont {J.}~\bibnamefont
  {{Ye}}},\ }\bibfield  {title} {\enquote {\bibinfo {title} {{Gapless
  spin-fluid ground state in a random quantum Heisenberg magnet}},}\ }\href
  {\doibase 10.1103/PhysRevLett.70.3339} {\bibfield  {journal} {\bibinfo
  {journal} {Phys. Rev. Lett.}\ }\textbf {\bibinfo {volume} {70}},\ \bibinfo
  {pages} {3339} (\bibinfo {year} {1993})},\ \Eprint
  {http://arxiv.org/abs/cond-mat/9212030} {cond-mat/9212030} \BibitemShut
  {NoStop}%
\bibitem [{\citenamefont {{Kitaev}}(2015)}]{kitaev2015talk}%
  \BibitemOpen
  \bibfield  {author} {\bibinfo {author} {\bibfnamefont {A.~Y.}\ \bibnamefont
  {{Kitaev}}},\ }\bibfield  {title} {\enquote {\bibinfo {title} {{Talks at
  KITP, University of California, Santa Barbara}},}\ }\href
  {http://online.kitp.ucsb.edu/online/entangled15/} {\bibfield  {journal}
  {\bibinfo  {journal} {Entanglement in Strongly-Correlated Quantum Matter}\ }
  (\bibinfo {year} {2015})}\BibitemShut {NoStop}%
\bibitem [{\citenamefont {{Shankar}}(1994)}]{shankar_rg}%
  \BibitemOpen
  \bibfield  {author} {\bibinfo {author} {\bibfnamefont {R.}~\bibnamefont
  {{Shankar}}},\ }\bibfield  {title} {\enquote {\bibinfo {title}
  {{Renormalization-group approach to interacting fermions}},}\ }\href
  {\doibase 10.1103/RevModPhys.66.129} {\bibfield  {journal} {\bibinfo
  {journal} {Rev. Mod. Phys.}\ }\textbf {\bibinfo {volume} {66}},\ \bibinfo
  {pages} {129} (\bibinfo {year} {1994})},\ \Eprint
  {http://arxiv.org/abs/cond-mat/9307009} {arXiv:cond-mat/9307009 [cond-mat]}
  \BibitemShut {NoStop}%
\bibitem [{\citenamefont {{Sachdev}}(2015)}]{SS15}%
  \BibitemOpen
  \bibfield  {author} {\bibinfo {author} {\bibfnamefont {S.}~\bibnamefont
  {{Sachdev}}},\ }\bibfield  {title} {\enquote {\bibinfo {title}
  {{Bekenstein-Hawking Entropy and Strange Metals}},}\ }\href {\doibase
  10.1103/PhysRevX.5.041025} {\bibfield  {journal} {\bibinfo  {journal} {Phys.
  Rev. X}\ }\textbf {\bibinfo {volume} {5}},\ \bibinfo {eid} {041025} (\bibinfo
  {year} {2015})},\ \Eprint {http://arxiv.org/abs/1506.05111} {arXiv:1506.05111
  [hep-th]} \BibitemShut {NoStop}%
\bibitem [{\citenamefont {{Georges}}\ \emph {et~al.}(2001)\citenamefont
  {{Georges}}, \citenamefont {{Parcollet}},\ and\ \citenamefont
  {{Sachdev}}}]{GPS01}%
  \BibitemOpen
  \bibfield  {author} {\bibinfo {author} {\bibfnamefont {A.}~\bibnamefont
  {{Georges}}}, \bibinfo {author} {\bibfnamefont {O.}~\bibnamefont
  {{Parcollet}}}, \ and\ \bibinfo {author} {\bibfnamefont {S.}~\bibnamefont
  {{Sachdev}}},\ }\bibfield  {title} {\enquote {\bibinfo {title} {{Quantum
  fluctuations of a nearly critical Heisenberg spin glass}},}\ }\href {\doibase
  10.1103/PhysRevB.63.134406} {\bibfield  {journal} {\bibinfo  {journal} {Phys.
  Rev. B}\ }\textbf {\bibinfo {volume} {63}},\ \bibinfo {eid} {134406}
  (\bibinfo {year} {2001})},\ \Eprint {http://arxiv.org/abs/cond-mat/0009388}
  {arXiv:cond-mat/0009388 [cond-mat.str-el]} \BibitemShut {NoStop}%
\bibitem [{\citenamefont {Fu}\ and\ \citenamefont
  {Sachdev}(2016)}]{Fu:2016yrv}%
  \BibitemOpen
  \bibfield  {author} {\bibinfo {author} {\bibfnamefont {W.}~\bibnamefont
  {Fu}}\ and\ \bibinfo {author} {\bibfnamefont {S.}~\bibnamefont {Sachdev}},\
  }\bibfield  {title} {\enquote {\bibinfo {title} {{Numerical study of fermion
  and boson models with infinite-range random interactions}},}\ }\href
  {\doibase 10.1103/PhysRevB.94.035135} {\bibfield  {journal} {\bibinfo
  {journal} {Phys. Rev.}\ }\textbf {\bibinfo {volume} {B94}},\ \bibinfo {pages}
  {035135} (\bibinfo {year} {2016})},\ \Eprint
  {http://arxiv.org/abs/1603.05246} {arXiv:1603.05246 [cond-mat.str-el]}
  \BibitemShut {NoStop}%
\bibitem [{\citenamefont {Azeyanagi}\ \emph {et~al.}(2018)\citenamefont
  {Azeyanagi}, \citenamefont {Ferrari},\ and\ \citenamefont
  {Schaposnik~Massolo}}]{Azeyanagi2018}%
  \BibitemOpen
  \bibfield  {author} {\bibinfo {author} {\bibfnamefont {T.}~\bibnamefont
  {Azeyanagi}}, \bibinfo {author} {\bibfnamefont {F.}~\bibnamefont {Ferrari}},
  \ and\ \bibinfo {author} {\bibfnamefont {F.~I.}\ \bibnamefont
  {Schaposnik~Massolo}},\ }\bibfield  {title} {\enquote {\bibinfo {title}
  {{Phase Diagram of Planar Matrix Quantum Mechanics, Tensor, and
  Sachdev-Ye-Kitaev Models}},}\ }\href {\doibase
  10.1103/PhysRevLett.120.061602} {\bibfield  {journal} {\bibinfo  {journal}
  {Phys. Rev. Lett.}\ }\textbf {\bibinfo {volume} {120}},\ \bibinfo {pages}
  {061602} (\bibinfo {year} {2018})},\ \Eprint
  {http://arxiv.org/abs/1707.03431} {arXiv:1707.03431 [hep-th]} \BibitemShut
  {NoStop}%
\bibitem [{\citenamefont {Gu}\ \emph {et~al.}(2019)\citenamefont {Gu},
  \citenamefont {Kitaev}, \citenamefont {Sachdev},\ and\ \citenamefont
  {Tarnopolsky}}]{GKST19}%
  \BibitemOpen
  \bibfield  {author} {\bibinfo {author} {\bibfnamefont {Y.}~\bibnamefont
  {Gu}}, \bibinfo {author} {\bibfnamefont {A.}~\bibnamefont {Kitaev}}, \bibinfo
  {author} {\bibfnamefont {S.}~\bibnamefont {Sachdev}}, \ and\ \bibinfo
  {author} {\bibfnamefont {G.}~\bibnamefont {Tarnopolsky}},\ }\bibfield
  {title} {\enquote {\bibinfo {title} {{Notes on the complex Sachdev-Ye-Kitaev
  model}},}\ }\href@noop {} {\bibfield  {journal} {\bibinfo  {journal} {to
  appear}\ } (\bibinfo {year} {2019})}\BibitemShut {NoStop}%
\bibitem [{\citenamefont {Maldacena}\ and\ \citenamefont
  {Qi}(2018)}]{Maldacena:2018lmt}%
  \BibitemOpen
  \bibfield  {author} {\bibinfo {author} {\bibfnamefont {J.}~\bibnamefont
  {Maldacena}}\ and\ \bibinfo {author} {\bibfnamefont {X.-L.}\ \bibnamefont
  {Qi}},\ }\bibfield  {title} {\enquote {\bibinfo {title} {{Eternal traversable
  wormhole}},}\ }\href@noop {} {\  (\bibinfo {year} {2018})},\ \Eprint
  {http://arxiv.org/abs/1804.00491} {arXiv:1804.00491 [hep-th]} \BibitemShut
  {NoStop}%
\bibitem [{\citenamefont {Varma}\ \emph {et~al.}(1989)\citenamefont {Varma},
  \citenamefont {Littlewood}, \citenamefont {Schmitt-Rink}, \citenamefont
  {Abrahams},\ and\ \citenamefont {Ruckenstein}}]{Varma89}%
  \BibitemOpen
  \bibfield  {author} {\bibinfo {author} {\bibfnamefont {C.~M.}\ \bibnamefont
  {Varma}}, \bibinfo {author} {\bibfnamefont {P.~B.}\ \bibnamefont
  {Littlewood}}, \bibinfo {author} {\bibfnamefont {S.}~\bibnamefont
  {Schmitt-Rink}}, \bibinfo {author} {\bibfnamefont {E.}~\bibnamefont
  {Abrahams}}, \ and\ \bibinfo {author} {\bibfnamefont {A.~E.}\ \bibnamefont
  {Ruckenstein}},\ }\bibfield  {title} {\enquote {\bibinfo {title}
  {{Phenomenology of the normal state of Cu-O high-temperature
  superconductors}},}\ }\href {\doibase 10.1103/PhysRevLett.63.1996} {\bibfield
   {journal} {\bibinfo  {journal} {Phys. Rev. Lett.}\ }\textbf {\bibinfo
  {volume} {63}},\ \bibinfo {pages} {1996} (\bibinfo {year}
  {1989})}\BibitemShut {NoStop}%
\bibitem [{\citenamefont {Faulkner}\ \emph {et~al.}(2013)\citenamefont
  {Faulkner}, \citenamefont {Iqbal}, \citenamefont {Liu}, \citenamefont
  {McGreevy},\ and\ \citenamefont {Vegh}}]{Faulkner13}%
  \BibitemOpen
  \bibfield  {author} {\bibinfo {author} {\bibfnamefont {T.}~\bibnamefont
  {Faulkner}}, \bibinfo {author} {\bibfnamefont {N.}~\bibnamefont {Iqbal}},
  \bibinfo {author} {\bibfnamefont {H.}~\bibnamefont {Liu}}, \bibinfo {author}
  {\bibfnamefont {J.}~\bibnamefont {McGreevy}}, \ and\ \bibinfo {author}
  {\bibfnamefont {D.}~\bibnamefont {Vegh}},\ }\bibfield  {title} {\enquote
  {\bibinfo {title} {{Charge transport by holographic Fermi surfaces}},}\
  }\href {\doibase 10.1103/PhysRevD.88.045016} {\bibfield  {journal} {\bibinfo
  {journal} {Phys. Rev.}\ }\textbf {\bibinfo {volume} {D88}},\ \bibinfo {pages}
  {045016} (\bibinfo {year} {2013})},\ \Eprint {http://arxiv.org/abs/1306.6396}
  {arXiv:1306.6396 [hep-th]} \BibitemShut {NoStop}%
\bibitem [{\citenamefont {Sachdev}(2010{\natexlab{a}})}]{SS10b}%
  \BibitemOpen
  \bibfield  {author} {\bibinfo {author} {\bibfnamefont {S.}~\bibnamefont
  {Sachdev}},\ }\bibfield  {title} {\enquote {\bibinfo {title} {{Strange metals
  and the AdS/CFT correspondence}},}\ }\href {\doibase
  10.1088/1742-5468/2010/11/P11022} {\bibfield  {journal} {\bibinfo  {journal}
  {J. Stat. Mech.}\ }\textbf {\bibinfo {volume} {1011}},\ \bibinfo {pages}
  {P11022} (\bibinfo {year} {2010}{\natexlab{a}})},\ \Eprint
  {http://arxiv.org/abs/1010.0682} {arXiv:1010.0682 [cond-mat.str-el]}
  \BibitemShut {NoStop}%
\bibitem [{\citenamefont {Faulkner}\ \emph {et~al.}(2011)\citenamefont
  {Faulkner}, \citenamefont {Liu}, \citenamefont {McGreevy},\ and\
  \citenamefont {Vegh}}]{Faulkner09}%
  \BibitemOpen
  \bibfield  {author} {\bibinfo {author} {\bibfnamefont {T.}~\bibnamefont
  {Faulkner}}, \bibinfo {author} {\bibfnamefont {H.}~\bibnamefont {Liu}},
  \bibinfo {author} {\bibfnamefont {J.}~\bibnamefont {McGreevy}}, \ and\
  \bibinfo {author} {\bibfnamefont {D.}~\bibnamefont {Vegh}},\ }\bibfield
  {title} {\enquote {\bibinfo {title} {{Emergent quantum criticality, Fermi
  surfaces, and AdS$_2$}},}\ }\href {\doibase 10.1103/PhysRevD.83.125002}
  {\bibfield  {journal} {\bibinfo  {journal} {Phys. Rev. D}\ }\textbf {\bibinfo
  {volume} {83}},\ \bibinfo {pages} {125002} (\bibinfo {year} {2011})},\
  \Eprint {http://arxiv.org/abs/0907.2694} {arXiv:0907.2694 [hep-th]}
  \BibitemShut {NoStop}%
\bibitem [{\citenamefont {Cubrovic}\ \emph {et~al.}(2009)\citenamefont
  {Cubrovic}, \citenamefont {Zaanen},\ and\ \citenamefont
  {Schalm}}]{Cubrovic:2009ye}%
  \BibitemOpen
  \bibfield  {author} {\bibinfo {author} {\bibfnamefont {M.}~\bibnamefont
  {Cubrovic}}, \bibinfo {author} {\bibfnamefont {J.}~\bibnamefont {Zaanen}}, \
  and\ \bibinfo {author} {\bibfnamefont {K.}~\bibnamefont {Schalm}},\
  }\bibfield  {title} {\enquote {\bibinfo {title} {{String Theory, Quantum
  Phase Transitions and the Emergent Fermi-Liquid}},}\ }\href {\doibase
  10.1126/science.1174962} {\bibfield  {journal} {\bibinfo  {journal}
  {Science}\ }\textbf {\bibinfo {volume} {325}},\ \bibinfo {pages} {439}
  (\bibinfo {year} {2009})},\ \Eprint {http://arxiv.org/abs/0904.1993}
  {arXiv:0904.1993 [hep-th]} \BibitemShut {NoStop}%
\bibitem [{\citenamefont {Sachdev}(2010{\natexlab{b}})}]{SS10}%
  \BibitemOpen
  \bibfield  {author} {\bibinfo {author} {\bibfnamefont {S.}~\bibnamefont
  {Sachdev}},\ }\bibfield  {title} {\enquote {\bibinfo {title} {{Holographic
  metals and the fractionalized Fermi liquid}},}\ }\href {\doibase
  10.1103/PhysRevLett.105.151602} {\bibfield  {journal} {\bibinfo  {journal}
  {Phys. Rev. Lett.}\ }\textbf {\bibinfo {volume} {105}},\ \bibinfo {pages}
  {151602} (\bibinfo {year} {2010}{\natexlab{b}})},\ \Eprint
  {http://arxiv.org/abs/1006.3794} {arXiv:1006.3794 [hep-th]} \BibitemShut
  {NoStop}%
\bibitem [{\citenamefont {Lee}\ \emph {et~al.}(2019{\natexlab{a}})\citenamefont
  {Lee}, \citenamefont {Patel}, \citenamefont {Trivedi},\ and\ \citenamefont
  {Sachdev}}]{Trivedi19}%
  \BibitemOpen
  \bibfield  {author} {\bibinfo {author} {\bibfnamefont {K.}~\bibnamefont
  {Lee}}, \bibinfo {author} {\bibfnamefont {A.~A.}\ \bibnamefont {Patel}},
  \bibinfo {author} {\bibfnamefont {N.}~\bibnamefont {Trivedi}}, \ and\
  \bibinfo {author} {\bibfnamefont {S.}~\bibnamefont {Sachdev}},\ }\bibfield
  {title} {\enquote {\bibinfo {title} {{Emergent interacting two-fluids in a
  disordered Hubbard model}},}\ }\href
  {http://meetings.aps.org/Meeting/MAR19/Session/H06.5} {\bibfield  {journal}
  {\bibinfo  {journal} {Bulletin of the Americal Physical Society}\ }\textbf
  {\bibinfo {volume} {64}},\ \bibinfo {pages} {H06.00005} (\bibinfo {year}
  {2019}{\natexlab{a}})}\BibitemShut {NoStop}%
\bibitem [{\citenamefont {Lee}\ \emph {et~al.}(2019{\natexlab{b}})\citenamefont
  {Lee}, \citenamefont {Patel}, \citenamefont {Sachdev},\ and\ \citenamefont
  {Trivedi}}]{Trivedi19a}%
  \BibitemOpen
  \bibfield  {author} {\bibinfo {author} {\bibfnamefont {K.}~\bibnamefont
  {Lee}}, \bibinfo {author} {\bibfnamefont {A.~A.}\ \bibnamefont {Patel}},
  \bibinfo {author} {\bibfnamefont {S.}~\bibnamefont {Sachdev}}, \ and\
  \bibinfo {author} {\bibfnamefont {N.}~\bibnamefont {Trivedi}},\ }\href@noop
  {} {\bibfield  {journal} {\bibinfo  {journal} {to appear}\ } (\bibinfo {year}
  {2019}{\natexlab{b}})}\BibitemShut {NoStop}%
\bibitem [{\citenamefont {{Davison}}\ \emph {et~al.}(2017)\citenamefont
  {{Davison}}, \citenamefont {{Fu}}, \citenamefont {{Georges}}, \citenamefont
  {{Gu}}, \citenamefont {{Jensen}},\ and\ \citenamefont
  {{Sachdev}}}]{Davison17}%
  \BibitemOpen
  \bibfield  {author} {\bibinfo {author} {\bibfnamefont {R.~A.}\ \bibnamefont
  {{Davison}}}, \bibinfo {author} {\bibfnamefont {W.}~\bibnamefont {{Fu}}},
  \bibinfo {author} {\bibfnamefont {A.}~\bibnamefont {{Georges}}}, \bibinfo
  {author} {\bibfnamefont {Y.}~\bibnamefont {{Gu}}}, \bibinfo {author}
  {\bibfnamefont {K.}~\bibnamefont {{Jensen}}}, \ and\ \bibinfo {author}
  {\bibfnamefont {S.}~\bibnamefont {{Sachdev}}},\ }\bibfield  {title} {\enquote
  {\bibinfo {title} {{Thermoelectric transport in disordered metals without
  quasiparticles: The Sachdev-Ye-Kitaev models and holography}},}\ }\href
  {\doibase 10.1103/PhysRevB.95.155131} {\bibfield  {journal} {\bibinfo
  {journal} {Phys. Rev. B}\ }\textbf {\bibinfo {volume} {95}},\ \bibinfo {eid}
  {155131} (\bibinfo {year} {2017})},\ \Eprint
  {http://arxiv.org/abs/1612.00849} {arXiv:1612.00849 [cond-mat.str-el]}
  \BibitemShut {NoStop}%
\end{thebibliography}%

\widetext
\appendix
\renewcommand{\theequation}{S\arabic{equation}}
\setcounter{equation}{0}
\renewcommand\thefigure{S\arabic{figure}}
\setcounter{figure}{0} 

\section{Supplementary Information}

\subsection{Real-space action and saddle point}

The Hamiltonian for our lattice model in real ($x$) space is given by
\beq
H = -t\sum_{\langle xx'\rangle}\sum_\alpha\left(c^\dagger_{x\alpha}c_{x'\alpha}+\mathrm{H.c}\right) - E_F\sum_{x}\sum_\alpha c^\dagger_{x\alpha}c_{x\alpha}+\frac{1}{(2N)^{3/2}}\sum_{\alpha_a}\sum_{x_a} U_{\alpha_a}(x_a)c^\dagger_{x_1\alpha_1}c^\dagger_{x_2\alpha_2}c_{x_3\alpha_3}c_{x_4\alpha_4},
\eeq
with $U_{\alpha_a}(x_a)$ defined in (\ref{current}). Upon averaging over the disorder using (\ref{UX1}) and introducing auxiliary fields $G,\Sigma$ to fix the values of the fermion two-point functions in the large-$N$ limit \cite{SS15,Balents2017,Chowdhury2018}, we obtain the action 
\begin{align}
&S = \int_0^\beta d\tau\left[\sum_x\sum_\alpha c^\dagger_{x\alpha}(\tau)\left(\partial_\tau-E_F\right)c_{x\alpha}(\tau) -t\sum_{\langle xx'\rangle}\sum_\alpha\left(c^\dagger_{x\alpha}(\tau)c_{x'\alpha}(\tau)+\mathrm{H.c}\right)\right] \nn
&-\frac{NU^2}{2}\int_0^\beta d\tau d\tau'\left[\sum_{x_1,x'_{a'}}\mathcal{F}(x_1,x'_{a'})G(x_1-x'_4,\tau-\tau')G(x_1-x'_3,\tau-\tau')G(x'_2-x_1,\tau'-\tau)G(x'_1-x_1,\tau'-\tau)\right] \nn
&-N\sum_{x,x'}\int_0^\beta d\tau d\tau' \Sigma(x-x',\tau-\tau')\left[G(x'-x,\tau'-\tau)-\frac{1}{N}\sum_\alpha c_{x'\alpha}(\tau')c^\dagger_{x\alpha}(\tau)\right],
\label{RSA}
\end{align}
where we have chosen a physically motivated translationally invariant form for $G,\Sigma$ after the disorder average, and $\mathcal{F}$ is defined in (\ref{UX2}). In the large-$N$ limit, the equations obtained by setting the variations $\delta S/\delta\Sigma=\delta S/\delta G=0$ after transforming the action to momentum space produce the Dyson equations (\ref{SYK1},\ref{SYK2}). The form of the interaction in real space is controlled by (\ref{UX1},\ref{UX2}), which doesn't have a simple expression for general $\epsilon_k$ and $\Xi,\Lambda$. However, for $\epsilon_k=k^2/(2m^\ast)-E_F$ and $\Lambda\rightarrow0$, we can express it as (focusing on $d=2$, with a bandwidth $W$)
\begin{align}
\mathcal{F}(x_1,x'_{a'})\Bigg|_{\Lambda=0,d=2} = \frac{m^{\ast 3}}{2}\int_0^{W m^\ast}\frac{dz}{(2\pi)^4}J_0\left(|x'_1-x_1|\sqrt{2z}\right)J_0\left(|x'_2-x_1|\sqrt{2z}\right)J_0\left(|x'_3-x_1|\sqrt{2z}\right)J_0\left(|x'_4-x_1|\sqrt{2z}\right),
\end{align}
where $J_0$ is the zeroth Bessel function of the first kind. This function $\mathcal{F}$ is peaked around the regions where $|x'_1-x_1|\pm|x'_2-x_1|\pm|x'_3-x_1|\pm|x'_4-x_1|=0$ and decays as $1/\sqrt{|x'_1-x_1||x'_2-x_1||x'_3-x_1||x'_4-x_1|}$ at large distances. Functions with similar characteristics can be obtained numerically for other values of $\epsilon_k$, $\Lambda$, $\Xi$ and $d$.  

\subsection{IR analysis}

To show the self-consistency of (\ref{GPlanckian}), we insert it into (\ref{SYK3}), and use (\ref{fancyC}). This yields
\begin{align}
&\Sigma(k,0\le\tau<\beta) = \left(\frac{T}{\sin(\pi T\tau)}\right)^{3/2}e^{-2\pi T\tau\mathbbm{C}\epsilon_k/U_\mathrm{eff}}\times\frac{U_\mathrm{eff}^{1/2}}{\Lambda^2}\int_{-\infty}^{\infty}d\epsilon k_2d\epsilon k_3\Xi\left(|\epsilon_{k_2}-\epsilon_{k_3}|\right)\Xi\left(|\epsilon_{k_3}-\epsilon_k|\right)\Xi(|\epsilon_{k_2}+\epsilon_{k_3}-2\epsilon_{k}|)\nn
&\times\mathcal{A}\left(\mathbbm{C}\frac{\epsilon_{k_2}}{U_\mathrm{eff}}\right)\mathcal{A}\left(\mathbbm{C}\frac{\epsilon_{k_3}}{U_\mathrm{eff}}\right)\mathcal{A}\left(\mathbbm{C}\frac{\epsilon_{k_2}+\epsilon_{k_3}-\epsilon_k}{U_\mathrm{eff}}\right)e^{-2\pi\mathbbm{C}(\epsilon_{k_2}+\epsilon_{k_3}-\epsilon_k)/U_\mathrm{eff}}.
\end{align}
Importantly, the $\tau$ dependence of the above is associated only with $\epsilon_k$ (and not $\epsilon_{k_2},\epsilon_{k_3}$) due to the resonance condition and the linearity of $\mathcal{E}_k$ in $\epsilon_k$ in the IR. Now, applying (\ref{SYK1}) leads to two conditions: ({\it i\/}) $\mathrm{Re}[\Sigma(k,\omega\rightarrow0)]+\epsilon_k = 0$, which we assume holds in the IR, and will be discussed further in the next subsection, and ({\it ii\/}) $\mathrm{Im}[G(k,\omega\rightarrow0)]=1/\mathrm{Im}[\Sigma(k,\omega\rightarrow0)]$, which yields
\begin{align}
&\mathcal{A}\left(\mathbbm{C}\frac{\epsilon_k}{U_\mathrm{eff}}\right)\int_{-\infty}^{\infty}d\epsilon k_2d\epsilon k_3\Xi\left(|\epsilon_{k_2}-\epsilon_{k_3}|\right)\Xi\left(|\epsilon_{k_3}-\epsilon_k|\right)\Xi(|\epsilon_{k_2}+\epsilon_{k_3}-2\epsilon_{k}|)\mathcal{A}\left(\mathbbm{C}\frac{\epsilon_{k_2}}{U_\mathrm{eff}}\right)\nn
&\times\mathcal{A}\left(\mathbbm{C}\frac{\epsilon_{k_3}}{U_\mathrm{eff}}\right)\mathcal{A}\left(\mathbbm{C}\frac{\epsilon_{k_2}+\epsilon_{k_3}-\epsilon_k}{U_\mathrm{eff}}\right)e^{-2\pi\mathbbm{C}(\epsilon_{k_2}+\epsilon_{k_3})/U_\mathrm{eff}}=\frac{\pi\Lambda^2}{4}\mathrm{sech}\left(2\pi\mathbbm{C}\frac{\epsilon_k}{U_\mathrm{eff}}\right).
\label{IA}
\end{align}
After fixing the value of $\mathbbm{C}$, this integral equation for $\mathcal{A}$ can be solved numerically by an iterative update scheme similar to that used for solving the Dyson equations for the SYK models. The value of $\mathbbm{C}$ must be determined by matching to the UV, which requires the direct numerical solution of the full Dyson equations, and is a function of $\Lambda/U_\mathrm{eff}$.  The sign of $\mathcal{A}$ is fixed to be $-1$ by noting that $G(k,0\le\tau<\beta)<0$. Since $\Xi(|\epsilon_k|)$ only depends on $|\epsilon_k|/\Lambda$, we can further see that $\mathcal{A}$ carries an implicit $\Lambda/U_\mathrm{eff}$ dependence. In the $\Lambda\rightarrow0$ limit, we obtain (\ref{AEk}). In Fig. \ref{GATest} we compare the exact numerical solutions of (\ref{SYK1},\ref{SYK3}) to the Ansatz determined by (\ref{GPlanckian}) and (\ref{IA}), and show that there is good agreement.
\begin{figure}
    \centering
    \includegraphics[width=0.96\textwidth]{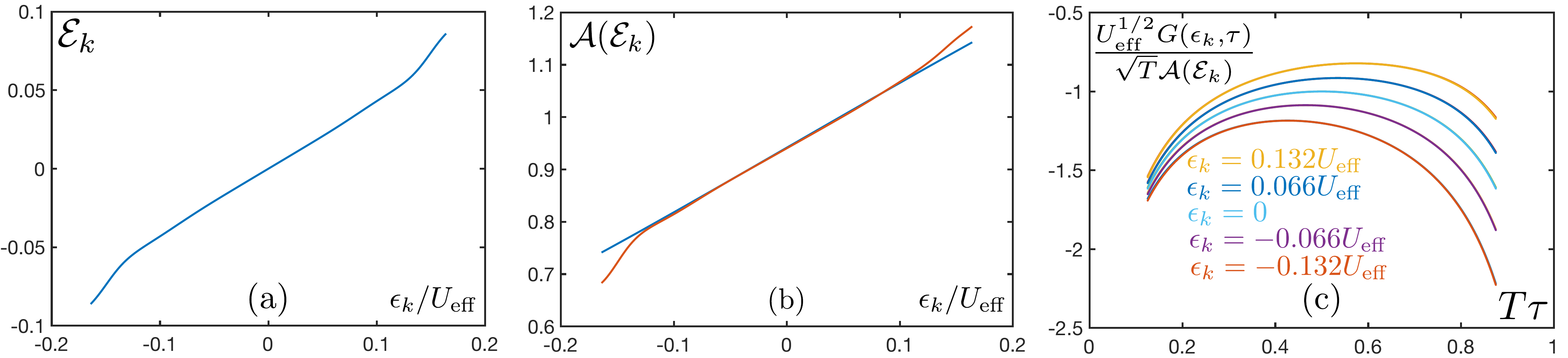}
    \caption{(a) Numerically determined $\mathcal{E}_k=-(1/(4\pi T)))d(G(\epsilon_k,\tau)/G(\epsilon,k,\beta-\tau))/d\tau|_{\tau=\beta/2}$. We find $\mathcal{E}_k\approx0.42\epsilon_k/U_\mathrm{eff}$. For $\epsilon_k$ outside the plot range, $G(\epsilon_k,\tau)$ starts to enter the crossover region (Fig. \ref{spectral} (a)), and $\mathcal{E}_k$ cannot be extracted. (b) Comparison of the numerical solution of (\ref{IA}) with $\mathbbm{C}=0.42$ (blue) to the numerically determined $\mathcal{A}(\mathcal{E}_k)=U^{1/2}_\mathrm{eff}e^{\pi\mathcal{E}_k}G(\epsilon,k,\tau=\beta/2)/\sqrt{T}$ (orange), with $\mathcal{E}_k$ determined as in (a) Good agreement is maintained outside the crossover region. (c) Comparison of the $\tau$ dependence of $G(\epsilon_k,\tau)$ between the numerical solution of (\ref{SYK1},\ref{SYK3}) (with $\mathcal{A}(\epsilon_k)$ determined as in (b)), and (\ref{GPlanckian}) (with $\mathcal{E}_k$ determined as in (a)). Each curve is actually two lines corresponding to the two quantities, which lie almost perfectly on top of each other.  We used $2\Lambda=0.16U_\mathrm{eff}$ and $T=0.0025U_\mathrm{eff}$, with the same function $\Xi$ as in Figs. \ref{remnant}, \ref{spectral}.}
    \label{GATest}
\end{figure}

\subsection{Transition/crossover at large $\epsilon/U$ in SYK}

Here, we review the solution of the Dyson equations for the SYK model, and explain the reason for the transition to a nearly-free fermion regime at large enough $\epsilon/U$. The Hamiltonian and Dyson equations for the SYK model are
\beq
H = \epsilon\sum_\alpha c^\dagger_\alpha c_\alpha + \frac{1}{(2N)^{3/2}}\sum_{\alpha_a}U_{\alpha_a}c^\dagger_{\alpha_1}c^\dagger_{\alpha_2}c_{\alpha_3}c_{\alpha_4},~~\Sigma(\tau) = -U^2G^2(\tau)G(-\tau),~~G^{-1}(\omega)=i\omega-\epsilon-\Sigma(\omega).
\eeq
At $T=0$, if we insert the conformal-limit solution \cite{SS15}
\beq
G(\tau>0) = -\frac{\cosh^{1/4}(2\pi\mathcal{E})}{U^{1/2}\pi^{1/4}\sqrt{1+e^{-4\pi\mathcal{E}}}}\frac{1}{\sqrt{\tau}},~~G(\tau<0)=-G(-\tau>0)e^{-2\pi\mathcal{E}},
\label{GSYK}
\eeq
into the Dyson equations, we get
\beq
\Sigma(\omega) = \frac{e^{\frac{\pi\mathcal{E}}{2}}\left(2-2 e^{2\pi\mathcal{E}}\right)\sqrt{U\eta_\mathrm{UV}}}{(2\pi)^{3/4}\left(e^{4\pi\mathcal{E}}+1\right)^{3/4}} + \frac{(1-i)e^{-4\pi\mathcal{E}}\left(e^{2\pi\mathcal{E}}-i\right) \cosh^{\frac{3}{4}}(2\pi\mathcal{E})}{\pi^{1/4}\left(e^{-4\pi\mathcal{E}}+1\right)^{3/2}}\sqrt{2U\omega},
\label{SSYK}
\eeq
where $\eta^{-1}_\mathrm{UV}\ll \omega^{-1}$ is a short-time UV cutoff that cuts off the short-time divergence in the Fourier transform of $\Sigma(\tau)$ to $\Sigma(\omega)$. Because $G(\omega)\sim1/(i\omega)$ at high frequencies, $\Sigma(\omega\rightarrow0)$ is dominated by contributions from the low frequency conformal-limit regime of $G$ (\ref{GSYK}). 

Demanding that $\mathrm{Re}[\Sigma(\omega\rightarrow0)]+\epsilon = 0$ (which is necessary for a self-consistent low-energy solution \cite{SY92}), we obtain the condition
\beq
\frac{e^{\frac{\pi\mathcal{E}}{2}}\left(2-2 e^{2\pi\mathcal{E}}\right)}{(2\pi)^{3/4}\left(e^{4\pi\mathcal{E}}+1\right)^{3/4}}=-\frac{\epsilon}{\sqrt{U\eta _\mathrm{UV}}},
\label{Fail1}
\eeq
The LHS of (\ref{Fail1}) is a bounded odd function of $\mathcal{E}$, ranging from $-0.263$ to $0.263$, so as long as $\eta_{UV}$ does not increase arbitrarily with $|\epsilon|$, (\ref{Fail1}) cannot hold for arbitrarily large $|\epsilon|$. 

To show that $\eta_\mathrm{UV}$ cannot be arbitrarily large, we note that $\eta_\mathrm{UV}$ should roughly be the scale where $|\omega|\sim|\epsilon+\Sigma(\omega)|$; for $|\omega|>\eta_\mathrm{UV}$, the $i\omega$ term in $G^{-1}(\omega)$ starts to dominate and the Green's function becomes that of a free fermion. From (\ref{SSYK}), we see that
\beq
\eta_\mathrm{UV} \sim \frac{U}{\cosh^{1/2}(2\pi\mathcal{E})},
\eeq
which is indeed bounded. Therefore the conformal-limit solution (\ref{GSYK}) cannot hold for arbitrarily large $|\epsilon|$, and must fail for $|\epsilon|\sim\mathcal{O}(U)$. Similar arguments also indicate the failure of (\ref{GPlanckian}) at large enough $|\epsilon_k|$ with the resonant Dyson equations (\ref{SYK1},\ref{SYK3}). It may be possible that a different set of Dyson equations with a different UV completion could allow the conformal-limit solutions (\ref{GSYK},\ref{GPlanckian}) to extend over a wider range of $\epsilon$ or $\epsilon_k$, but finding such a construction is beyond the scope of this letter.

\subsection{Other physical properties}

{\bf Specific heat.} The free energy density in our model at the large-$N$ saddle point can be derived by integrating out the fermions $c,c^\dagger$ in (\ref{RSA}) and substituting in the relation (\ref{SYK2}) between $\Sigma$ and $G$ \cite{Davison17}. It is given by
\beq
F = -N\int_k\left[T\sum_\omega\ln G^{-1}(k,\omega) -\frac{3}{4}T\sum_\omega\Sigma(k,\omega)G(k,\omega)\right],
\eeq
which can be further reduced to an integral over the dispersion $\epsilon_k$ by linearizing around the RFS. In our $T\ll U\ll E_F,W$ approximation, there is an effective particle-hole symmetry about the RFS, and the specific heat is then simply given by $C_V=-T\partial^2F/\partial T^2$. We can then express the specific heat as
\beq
C_V = N\int_k C_V(k) = N\left(\oint_\mathrm{RFS}\frac{d^{d-1}a_k}{(2\pi)^d}\frac{1}{|\nabla_k\epsilon_k|}\right)\int_{-\infty}^{\infty}d\epsilon_k ~C_V(\epsilon_k) .
\eeq
Since the $T,\tau,U_\mathrm{eff}$ dependencies of (\ref{GPlanckian}) are the same as that of the complex SYK model, we expect $C_V(\epsilon_k)=(T/U_\mathrm{eff})h(\epsilon_k/U_\mathrm{eff})$ as $T\rightarrow0$ \cite{Davison17}, and hence the interaction strength $U_\mathrm{eff}$ also cancels out in the total specific heat. For the case of $\Lambda\rightarrow0$, we found numerically that 
\beq
\int_{-\infty}^{\infty}d\epsilon_k ~C_V(\epsilon_k) \approx \int_{-\epsilon_k^\ast}^{\epsilon_k^\ast}d\epsilon_k ~C_V(\epsilon_k) \approx 0.50\times T,
\eeq
for $T\ll U_\mathrm{eff}$, where $\epsilon_k^\ast\approx 0.24\times U_\mathrm{eff}$ is the crossover scale to nearly-free fermions. For comparision, the textbook Fermi liquid specific heat is 
\beq
C_V^\mathrm{FL}=N\left(\oint_\mathrm{FS}\frac{d^{d-1}a_k}{(2\pi)^d}\frac{1}{|\nabla_k\epsilon_k|}\right)\frac{\pi^2}{3}T.
\eeq
Thus, our model has a $T$-linear specific heat proportional to the effective mass as $T\rightarrow0$, as in a Fermi liquid, but with a different numerical constant of proportionality. 

{\bf Lorenz number.} The Lorenz number $L$ is the ratio of the thermal ($\kappa$) and electrical ($\sigma$) conductivities, $L=\kappa/(\sigma T)$. Due to the effective particle-hole symmetry about the RFS in the large $E_F$ approximation, thermoelectric effects can be ignored, and $\kappa$ is then simply given by the two-point function of the energy current operator via the Kubo formula. The contribution $I^\mathrm{E}_\mathrm{I}$ to the energy current $I^\mathrm{E}$ that determines $\kappa$ is analogous to (\ref{thecurrent}); the term analogous to (\ref{current}) does not contribute to $\kappa$ for the same reason (\ref{current}) does not contribute to $\sigma$. We have
\beq
I^\mathrm{E}_\mathrm{I} = \int_k\sum_\alpha\mathbf{v}_kc^\dagger_{k\alpha}\partial_\tau c_{k\alpha},~~\kappa = \lim_{\Omega\rightarrow0}\frac{\mathrm{Im}[\langle I^\mathrm{E}_\mathrm{I}\cdot I^\mathrm{E}_\mathrm{I}\rangle^R(\Omega)]}{T\Omega d}.
\eeq
The Lorenz number is then given by \cite{Patel2017}
\beq
L = \frac{\int_{-\infty}^{\infty} d\epsilon~\int_{-\infty}^{\infty}d\omega~\left(\frac{\omega}{T}\right)^2\mathrm{sech}^2\left(\frac{\omega}{2T}\right)A^2(\epsilon,\omega)}{\int_{-\infty}^{\infty}d\epsilon~\int_{-\infty}^{\infty}d\omega~\mathrm{sech}^2\left(\frac{\omega}{2T}\right)A^2(\epsilon,\omega)}.
\eeq
Due to the $\omega/T$ scaling of $A$ near the RFS, we can see that $L$ will be independent of $T$ as $T\rightarrow0$ up to very small non-universal contributions coming from large $|\epsilon|>\epsilon_k^\ast$. In the special case of $\Lambda\rightarrow0$, cutting off the $\epsilon$ integrals at $\epsilon_k^\ast$ gives $L\approx 1.40\times(k_B/e)^2$. If we assume instead that the linearized expression for $\mathcal{E}_k$ in (\ref{fancyC}) holds for the full range of $\epsilon$ integration, then we obtain $L\approx 9.83\times(k_B/e)^2$. In a Fermi liquid $L=L_0=(\pi^2/3)(k_B/e)^2$ as $T\rightarrow0$.   

\end{document}